\def\fxo{$\ f_x/f_{o}\ $}

\def\ecs{ ${\rm erg~cm}^{-2}~{\rm s}^{-1}$ }    
\def\es{ ${\rm erg~\rm s}^{-1}$ }    
\def\cm{ ${\rm cm~\rm}^{-2}$ }    
\def\kms{ ${\rm km~\rm s}^{-1}$ }

\documentclass[useAMS,usenatbib]{mn2e}
\usepackage{graphicx}
\usepackage{txfonts}
\usepackage{natbib}

\usepackage[english]{babel}
\usepackage{color}
\DeclareGraphicsExtensions{.pdf,.ps,.jpg}
\usepackage{subfigure}

\def\aj{AJ}%
\def\araa{ARA\&A}%
\def\apj{ApJ}%
\def\apjl{ApJ}%
\def\apjs{ApJS}%
\def\aap{A\&A}%
\def\mnras{MNRAS}%
\def\memras{MmRAS}%
\def\pasp{PASP}%
\def\nat{Nature}%
\newcommand{\ltsima}{$\; \buildrel < \over \sim \;$}
\newcommand{\simlt}{\lower.5ex\hbox{\ltsima}} 
\newcommand{\gtsima}{$\; \buildrel > \over \sim \;$}
\newcommand{\simgt}{\lower.5ex\hbox{\gtsima}} 

\newcommand{\xmm}{{XMM-\emph{Newton}}}

\newcommand{\mum}{\:\mu\mbox{\scriptsize m}}

\title[Fx/Fo$>$50]{Exploring the Active Galactic Nuclei population with extreme X-ray to optical flux ratios (\fxo $>50$)}
\author[Della Ceca et al.]{R. Della Ceca$^{1}$\thanks{E-mail: roberto.dellaceca@brera.inaf.it},
 F.J. Carrera$^{2}$,
 A. Caccianiga$^{1}$,
 P. Severgnini$^{1}$,
 L. Ballo$^{1}$,
 V. Braito$^{1}$,
\newauthor
 A. Corral$^{3}$,
 A. Del Moro$^{4}$,
 S. Mateos$^{2}$,
 A. Ruiz$^{2}$, and 
 M.G. Watson$^{5}$\\
$^{1}$INAF-Osservatorio Astronomico di Brera, via Brera 28, 20121 Milano, Italy\\
$^{2}$Instituto de Fisica de Cantabria (CSIC-UC), Avenida de los Castros, E-39005 Santander, Spain\\
$^{3}$National Observatory of Athens (NOA), Palaia Penteli, GR-15236 Athens, Greece\\
$^{4}$Department of Physics, Durham University, South Road, Durham DH1 3LE, UK\\
$^{5}$Department of Physics \& Astronomy, University of Leicester, Leicester, LE1 7HR, UK
}

\begin{document}

\date{}

\pagerange{\pageref{firstpage}--\pageref{lastpage}} 

\maketitle

\label{firstpage}

\begin{abstract}

The cosmic history of the growth of supermassive black holes in galactic centers parallels that of star--formation in the Universe. However, an important fraction of this growth occurs inconspicuously in obscured objects, where ultraviolet/optical/near-infrared emission is heavily obscured by dust. Since the X--ray flux is less attenuated, a high X-ray--to--optical flux ratio (\fxo) is expected to be an efficient tool to find out these obscured accreting sources. We explore here via optical spectroscopy, X-ray spectroscopy and infrared photometry the most extreme cases of this population (those with \fxo $>50$, EXO50 sources hereafter), using a well defined sample of 7 X-ray sources extracted from the 2XMM catalogue. 
Five EXO50 sources ($\sim 70$ percent of the sample) in the bright flux regime explored by our survey ($f_{(2-10 keV)} \geq 1.5\times 10^{-13}$ \ecs) are associated with obscured AGN ($N_H > 10^{22}$ cm$^{-2}$), spanning a redshift range between 0.75 and 1 and characterised by 2-10 keV intrinsic luminosities in the QSO regime (e.g. well in excess to $10^{44}$ \es). We did not find compelling evidence of Compton Thick AGN. Overall the EXO50 Type 2 QSOs do not seem to be different from standard X-ray selected Type 2 QSOs in terms of nuclear absorption; a very high AGN/host galaxy ratio seems to play a major role in explaining their extreme properties. Interestingly 3 out of 5 EXO50 Type 2 QSO objects can be classified as Extreme Dust Obscured Galaxies 
(EDOGs, $f_{24 \mum}/f_{R} \geq 2000$), suggesting that a very high AGN/host ratios (along with the large amount of dust absorption) could be the natural explanation also for a part of  the EDOG population.
The remaining 2 EXO50 sources are classified as BL Lac objects, having rather extreme properties, and  which are good candidates for TeV emission. 
\end{abstract}


\section{Introduction}

The study of high-z obscured quasars (Type 2 QSOs: the high luminosity counterpart of Seyfert 2 galaxies) is one of the hot topics of current extragalactic astronomy. Their observed distributions (e.g. redshift, flux and absorption) and physical properties (e.g. bolometric luminosity, black hole mass, accretion rate), compared to those of unobscured QSOs, are key parameters to understand and to test the orientation-based Unified Schemes (\citealt{antonucci1993}) for Active Galacic Nuclei (AGN) and to constrain the contribution of QSOs to the X$-$ray Background (XRB: e.g. \citealt{gilli2007}; \citealt{treister2009}). Even more important, there are now increasing evidence that the formation and growth of galaxies and their nuclear supermassive black holes (SMBHs, $M_{BH} > 10^6\ M_{\odot}$) are intimately related; feedback from actively accreting SMBH, AGN, must play a fundamental role in regulating both star formation and accretion throughout galaxy's history 
(e.g. \citealt{silk1998}, \citealt{granato2004}, \citealt{dimatteo2005}, \citealt{croton2006}, 
\citealt{hopkins2008}, \citealt{menci2008}, \citealt{king2010}, \citealt{faucher2012}). Although the nature of this relationship is still poorly understood, there are hints that most of the SMBH accretion takes place during an obscured quasar phase. The infrared (IR), optical and X$-$ray spectral analysis of a large sample of obscured quasars probably represents one of the best methods to test a possible coevolution between massive galaxies and AGN activity (see e.g. \citealt{hopkins2006}) and to investigate if and how the AGN feedback can affect the galaxy evolution (see e.g. \citealt{bower2006}). Indeed, in these sources the properties of the host galaxy can be studied using the optical and near-IR data, where the absorbed AGN is supposed to contribute marginally, while the AGN properties can be investigated using the X-ray and mid-IR data where, conversely, the AGN emission is dominant (see \citealt{ballo2014} and references therein). 
Obscured QSOs, a rare class of objects, are thus expected to have large values of the X-ray-to-optical flux ratio.
Furthermore, since the dust extinction increases in the ultraviolet (UV) while the X-ray absorption goes in the opposite direction, i.e. strongly decreasing going towards the high energies, a redshift dependence is also expected ($\sim (1+z)^{3.6}$; \citealt{fiore2003}). However, in spite of the outstanding progress, obtained by using medium and deep X--ray surveys (see e.g. \citealt{brusa2010} and references therein), the weakness, both in the X-ray and in the optical band, of the selected sources with high values of the X-ray-to-optical flux ratio usually prevents from a detailed analysis of the individual objects (see e.g. \citealt{koekemoer2004}, \citealt{civano2005}).

To explore the most extreme examples of obscured QSOs in the bright flux regime, we have started a project focused on the very ``high" X-ray-to-optical flux ratio, \fxo{} (see Eq. 1), population, {namely those sources with \fxo{} $>$50} (more than 15 times the average values of unobscured broad line AGN, e.g. \citealt{caccianiga2004}, \citealt{dellaceca2004}, \citealt{civano2012}) and with $f_x>10^{-13}$\ecs.
In this way we should efficiently select the best candidates to be obscured QSOs; at the same time, the source brightness  ensures that the optical spectroscopic identification can be achieved
\footnote{At the chosen X-ray flux limit of $f_x \sim 10^{-13}$ \ecs a source with \fxo$\sim$ 50 (300) has an optical R magnitude of $\sim$23 ($\sim$25).} for sources with reasonably good quality X-ray data ($>$few hundred counts) to carry out a reliable X-ray spectral analysis.
Hereafter we will call these sources EXO50 -- for ``extreme X-ray to optical flux ratio". 
One example of such extreme obscured objects is XBS~J021642.3-043553 with \fxo $\sim$ 200, for which the presence of a Type 2 QSO at $z\sim 2$ was spectroscopically confirmed with VLT/FORS 
\citep{severgnini2006}. Using the redshift information and the spectral energy distribution (SED), we estimated a stellar mass of $\sim 10^{11}$ M$_{\odot}$ for the host galaxy, which supports the strong link between high-redshift massive galaxies and powerful obscured high-redshift QSOs. 
In spite of the importance of obscured QSOs in the cosmological context, only less than a dozen of such extreme \fxo sources have been found and studied so far with good X-ray and optical data (see e.g. \citealt{gandhi2006}, 
\citealt{severgnini2006}, \citealt{delmoro2009}, \citealt{campisi2009}, \citealt{brusa2010}
\citealt{brusa2015}, \citealt{perna2014}). 

Here we discuss the first results obtained from this project on a small, but statistically complete and representative, sub-sample of 7 EXO50 objects with $f_x \geq 1.5\times 10^{-13}$ \ecs. It is worth mentioning that other two interesting (and rare) classes of extragalactic sources are expected to show up in the high \fxo domain explored here, namely BL Lac objects and high redshift (i.e. z$>0.6$ in order to have the 4000 \AA\  break shifted at longer wavelengths than the R band filter) clusters of galaxies. Clusters of galaxies, however, are not expected to be represented in our sample since we have restricted our search only to the point-like X-ray sources, thus strongly minimising their possible selection. 

This paper is organized as follows:  in Section 2 we discuss the strategy used to define a statistically representative  sample of sources with \fxo{} $>$50. In Section 3 we present the data accumulated so far (in the radio, IR, optical, and X-ray energy ranges) on the 7 EXO50 objects discussed here, while their properties, source by source, are discussed in Section 4.
 In Section 5 we compare the broad band, from IR to X-rays,  properties of these EXO50 sources with other similar objects found in the literature.
Summary and conclusions are presented in Section 6.  Throughout this paper, we consider the cosmological model with
($H_o$,$\Omega_M$,$\Omega_{\lambda}$)=(70,0.3,0.7);  results from other papers have been rescaled to this 
cosmological framework. All the optical/IR magnitudes reported here are in the Vega system. 
Unless differently quoted,  X-ray luminosities are intrinsic (i.e. unabsorbed) luminosities in the rest-frame $2-10$ keV energy range. In this paper we use the term Type 1 and Type 2 AGN as broad or narrow line (FWHM of the permitted emission lines $<1500$ \kms)
AGN, irrespective to their intrinsic X-ray luminosity, while we use the term Type 1 QSO and Type 2 QSO (or obscured QSO) as broad or narrow AGN with an intrinsic 2-10 keV luminosity in excess to $10^{44}$ \es. 
Finally, errors are at 90\% confidence level for the X-ray spectral parameters derived using XSPEC (as usually done in X-ray astronomy) and 68\% for all the other quantities.    

\section{The EXO50 Sample}

In obscured QSOs the nuclear UV/optical emission is suppressed by dust obscuration 
(thus leaving only the galaxy component visible), whilst the nuclear X-ray flux (e.g. in the 2-10 keV energy range),
even if exhibiting signatures of photoelectric absorption,
is less attenuated
(see e.g. \citealt{fiore2003}). Obscured QSOs are therefore expected to display large values of X-ray to optical flux ratio, defined here as: 
\begin{equation}
f_x/f_{o} = f_x / (\Delta \lambda \times 2.15\times 10^{-9} \times 10^{-0.4\times R})
\end{equation}
where $f_x$ refers to the observed 2-10 keV flux (corrected for Galactic absorption), R is the observed optical magnitude in the R band ($\lambda \simeq 6410 $\ \AA\ ) and $\Delta \lambda \simeq 1568 $\ \AA\  (see \citealt{fukugita1995}). 

We define EXO50 the sources with \fxo$> 50$; these are a rare class of X-ray emitting objects, 
as typically AGN have \fxo = 1-10 (see e.g. \citealt{caccianiga2004}, \citealt{dellaceca2004}, \citealt{civano2012}). 
For instance, in the XMM-Newton Bright Survey (XBS, \citealt{dellaceca2004}, \citealt{caccianiga2008}), a complete sample of bright ($f_{(0.5-4.5 keV)} \gtrsim 7\times 10^{-14}$  \ecs) X-ray selected sources almost completely identified (spectroscopic identification level $\sim 98\%$), EXO50 objects represents only $\sim 0.5\%$ of the source population.

To construct our sample of EXO50 sources in the bright flux regime 
we used one of the largest well defined and complete X-ray source sample derived so far (discussed in \citealt{mateos2008}), based on the 2XMM source catalog
\footnote{See http://xmmssc-www.star.le.ac.uk/Catalogue/xcat\_public\_2XMM.html}
(\citealt{watson2009}). 

First we have considered the following selection criteria:

a) ``point-like" X-ray sources (parameters EP\_EXTENT and  EP\_EXTENT\_ML in the source catalogue equal to 0), in order to minimise the presence of clusters of galaxies which is another possible class of high \fxo sources;

b) north of -10 deg declination, in order to be accessible to optical investigation from  
the Italian {\it Telescopio Nazionale Galileo} (TNG), from the {\it Large Binocular Telescope} (LBT) and from the Spanish {\it Gran Telescopio Canarias} (GTC);

c) sources selected in the 2-10 keV energy band with a $f_{2-10\ keV}$ $> 10^{-13}$ \ecs;

d) only serendipitous sources (e.g. sources that are not related to nearby galaxies or to the XMM-Newton pointing) have been taken into consideration. We have also excluded the serendipitous sources already classified as 
non AGN.

Starting from a source list of 9431 sources (see \citealt{mateos2008}), these first selection criteria provide us with a list of about 600 X-ray emitting objects. 

The second step was to select the X-ray sources with \fxo$>$50, using:  

i) optical source archives (e.g. APM, Sloan Digital Sky Survey - SDSS, etc);

ii) optical imaging data from archives (STScI Digitized Sky Survey, ESO archive, etc..);

iii) imaging data from  our dedicated observing programs (e.g. TNG imaging).

We have looked at and visually inspected all the selected X-ray sources for possible optical counterparts, using a search radius of 4 arcsec (see \citealt{caccianiga2008}), and derived their R magnitude. For all the sources reported and discussed here (see Table 1), the offset between the X-ray and the optical position, derived {a posteriori}, 
is below 2.1 arcsec, fully consistent with the results obtained by  \cite{caccianiga2008} from the analysis of the X-ray to optical offset for the X-ray sources belonging to the XBS. 
The accumulated magnitudes, combined with the observed 2-10 keV fluxes, have been used to compute the \fxo ratio and thus to select EXO50 sources. 

In this paper we consider a first complete sample of EXO50 sources having $f_x$ $> 1.5\times 10^{-13}$ \ecs; this sample is composed by the 7 EXO50 objects reported in Table 1.
In Figure 1 we show the \fxo distribution of the sample composed by all the serendipitous point-like sources 
in the covered area having $f_x$ $> 1.5\times 10^{-13}$ \ecs (i.e. the starting sample, 310 objects). 
In the same figure we show how extreme the 7 sources are with respect to the starting sample; the EXO50 sources discussed here 
represent only $\sim$ 2\% of the population considered.
We stress that according to the selection criteria discussed above and the optical material at our disposal these 7 objects are the only EXO50 sources, north of -10 deg declination, 
with $f_x$ $> 1.5\times 10^{-13}$ \ecs in the sky survey area defined in \cite{mateos2008}; the sky coverage investigated to find out these 7 objects is about 60.4 sq deg.

\begin{figure}
\begin{center}
\resizebox{0.46\textwidth}{!}{
\rotatebox{0}{
\includegraphics{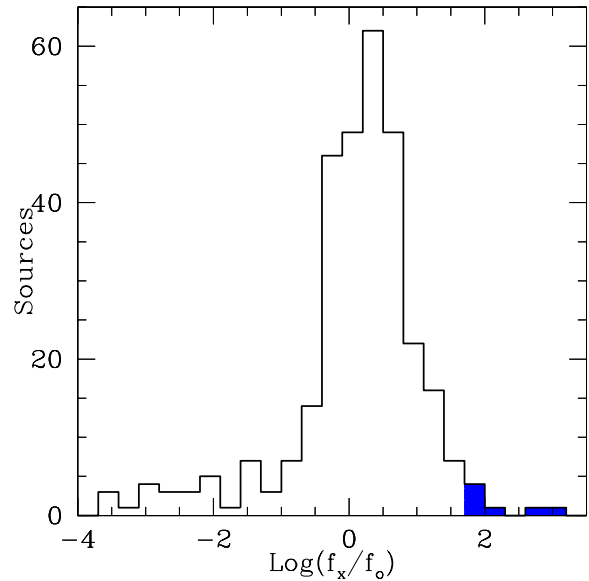}
}}
\caption{Histogram of the \fxo distribution of the starting sample with $f_x$ $> 1.5\times 10^{-13}$ \ecs (310 objects). 
We also show as filled histogram the \fxo distribution of the 7 EXO50 sources discussed in this paper.}
\label{fig}
\end{center}
\end{figure}

As detailed in the following sections, out of the 7 EXO50 sources 2 are BL Lac objects, 3 are confirmed Type 2 QSO and 2 remain spectroscopically unidentified, although their broad band properties strongly suggest an obscured QSO nature. 
Using the JAVA software GALAXYCOUNT (\citealt{ellis2007}) and considering the optical magnitudes of the proposed optical counterparts (see Table 1) we estimate less than 0.3 normal galaxies by chance in the total area covered from the 
7 error boxes, assuring us about the reliability of the spectroscopic identification proposed here.

\begin{table*} 
\begin{center}
\caption{The complete sample of EXO50 objects discussed in this paper 
\label{tab1}
}
 \begin{tabular}{rrrrrrrrrrr}
\hline
n & 2XMMName               & {\it f}$_x$      &  R            & \fxo  & ID     & z     & N$_H^a$                         & $\Gamma^a$                  & $L_x$     & $f_r$          \\
  &                & 10$^{-13}$ cgs &          &      &        &       & 10$^{22}$  cm $^{-2}$                 &                           & 10$^{45}$ cgs &          mJy      \\     
(1)              & (2)        & (3)      & (4)  & (5)    & (6)   & (7)                        & (8)                       & (9)       & (10)   & (11)         \\        
\hline
\hline
1 & J022256.9-024258        &  3.50      & 22.1$\pm$0.4  &  72   & AGN2   & 1.004 &  7.5$^{+3.2}_{-2.9}$       &  1.80$^{+0.4}_{-0.4}$     & 1.9      & 429$\pm$14        \\
2 & J100038.9+050955        &  1.63      & 23.6$\pm$0.4  & 138   & -      & -     & $>$0.4$^b$                 &  1.9$^c$                  & -        & -                  \\        
3 & J121026.5+392908        & 36.1       & 19.16$^d$$\pm$0.02  &  50   & BLLac  & 0.617 &  0.06$^{+0.004}_{-0.002}$  &  2.23$^{+0.01}_{-0.01}$   & 6.4      & 19.0$\pm$0.7       \\   
4 & J121134.2+390054$^e$    &  8.2       & 20.77$^d$$\pm$0.05  &  50   & BLLac  & 0.890 &  0.19$^{+0.04}_{-0.04}$    &  2.21$^{+0.07}_{-0.06}$   & 3.7      & 10.6$\pm$0.6       \\
5 & J123204.9+215254$^e$    & 10.0       & 23.9$\pm$0.4        & 1118  & AGN2   & 0.763 &  3.61$^{+1.03}_{-0.92}$    &  1.31$^{+0.25}_{-0.21}$   & 1.9      & -                    \\
6 & J135055.7+642857        &  1.5       & 25.0$\pm$0.8        & 458   & -      & -     & $>$0.1$^b$                 &  2.02$^{+0.13}_{-0.08}$   & -        & 183.5$\pm$5.5        \\
7 & J143623.8+631726        &  2.53      & 22.2$^d$$\pm$0.2  &  55  & AGN2    & 0.893 &  1.46$^{+0.39}_{-0.24}$    &  1.69$^{+0.15}_{-0.11}$   & 0.84     & -                    \\
\hline
\end{tabular}
\end{center}
Columns are as follows: 
1) Number used to mark the object in the plots shown in Section 5;
2) source name;
3) X-ray flux in the 2-10 keV energy band (MOS normalisation) corrected for Galactic absorption 
in units of 10$^{-13}$ \ecs;
4) R band magnitude;      
5) X-ray to optical flux ratio; 
6) optical spectroscopic classification;
7) redshift;
8) intrinsic absorption and 90\% confidence interval in units of 10$^{22}$ cm$^{-2}$ from fits to X-ray spectra;
9) power law photon index, $\Gamma$,  in the X-ray energy range, and 90\% confidence interval;
10) intrinsic X-ray luminosity in the 2-10 keV rest frame energy range, in units of 10$^{45}$ \es;
11) radio flux at 1.4 GHz in mJy.
Notes: $^a$ - For all the sources the best fit spectral model in the X-ray energy band is 
              a simple absorbed power-law model;
       $^b$ - The 90\% lower limit on N$_H$ obtained assuming z=0;
       $^c$ - This parameter has been fixed;
       $^d$ - For these sources the R band magnitude has been obtained from the SDSS r "model" magnitudes 
              assuming r-R = 0.27 (r$_{SDSS\ AB}$ = R$_{Vega}$ + r$_{AB}$(Vega) with r$_{AB}$(Vega)=0.27);   
       $^e$ - For these two sources we propose here a different redshift with respect to that already reported in the literature. See Section 4 for details.               
\end{table*}

\begin{figure*}
\label{fig}  
\centering
\subfigure{ 
  \includegraphics[width=6cm,angle=0]{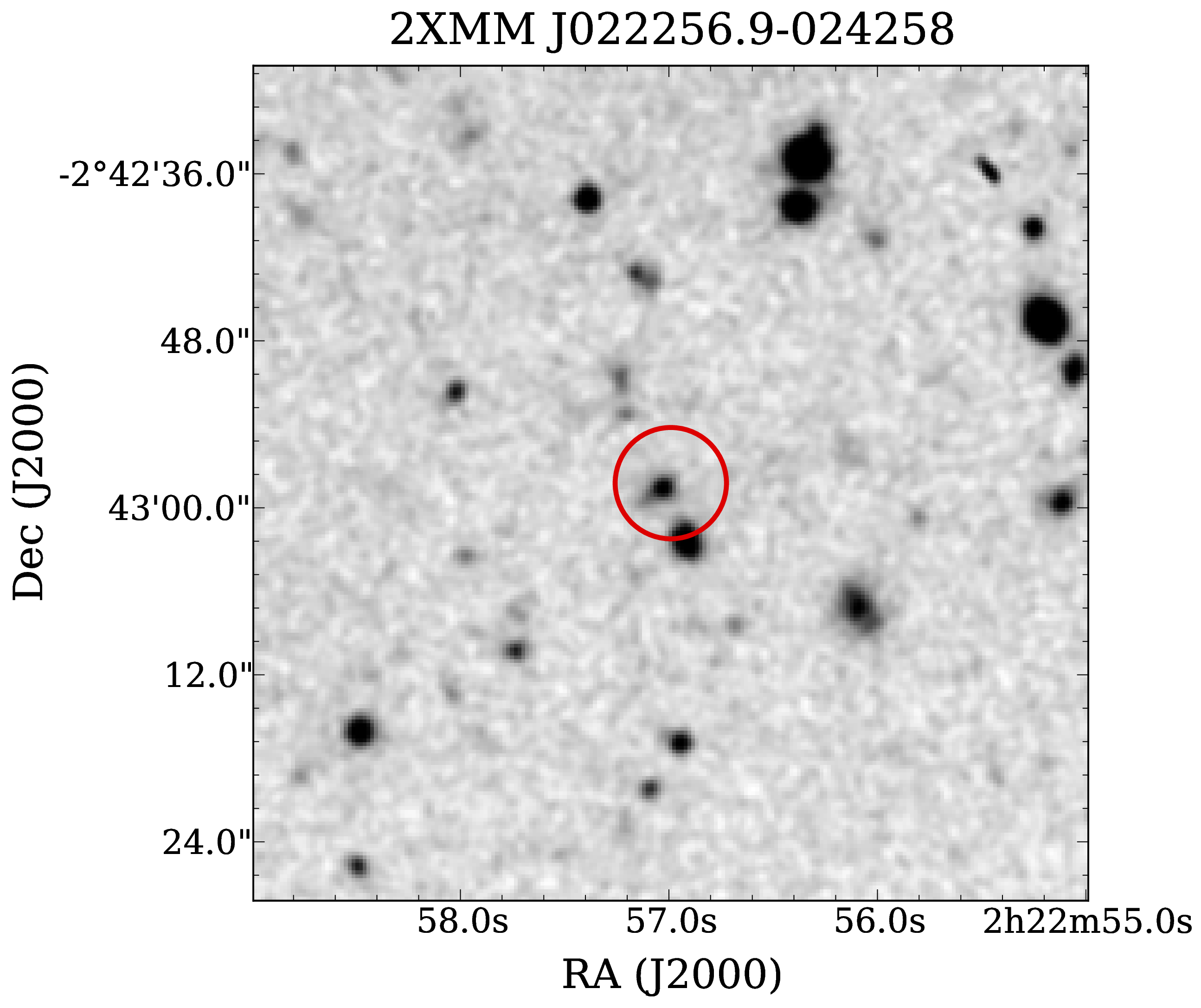}
  \includegraphics[width=6cm,angle=0]{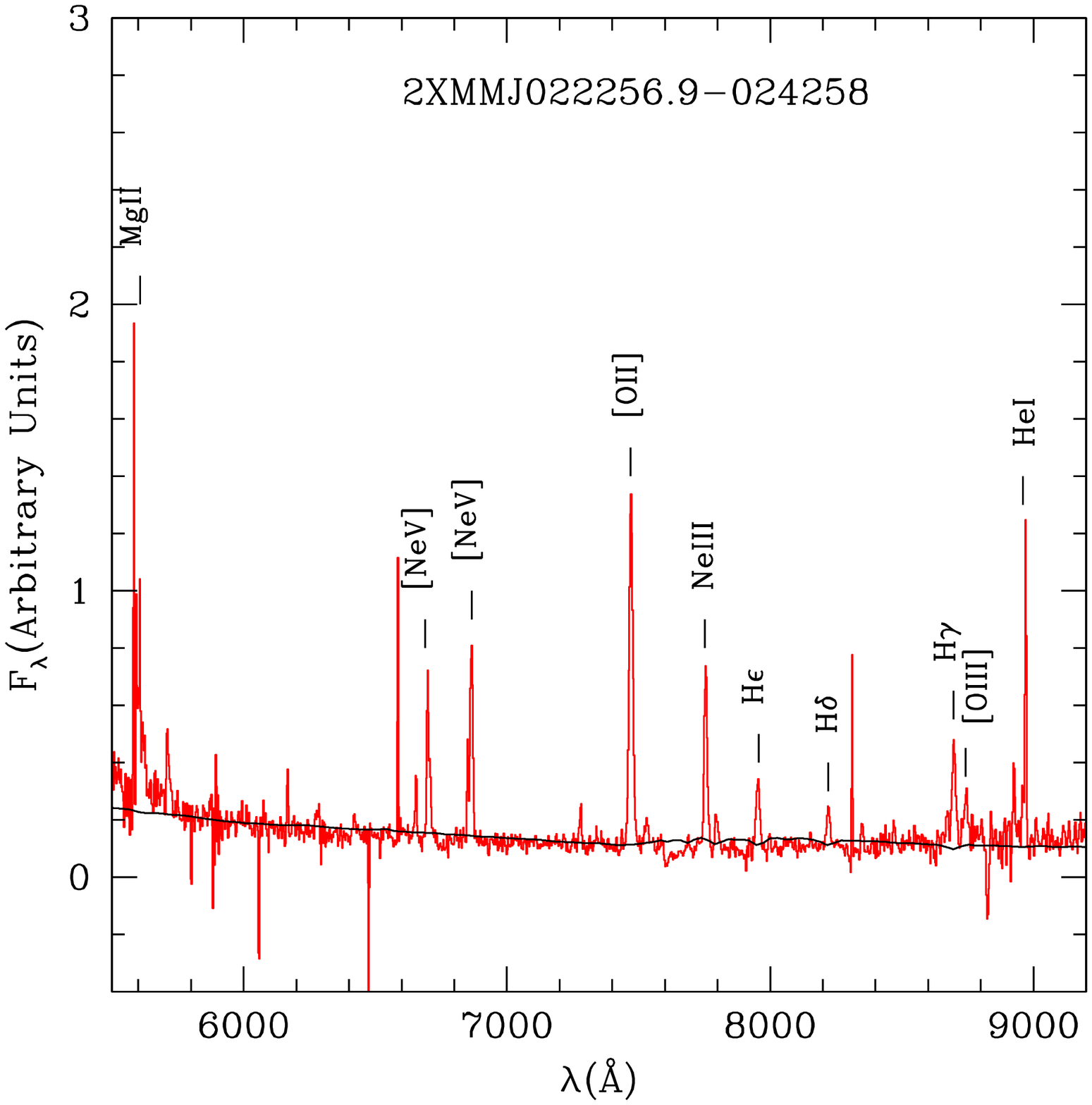}
  \includegraphics[width=6.5cm,angle=0]{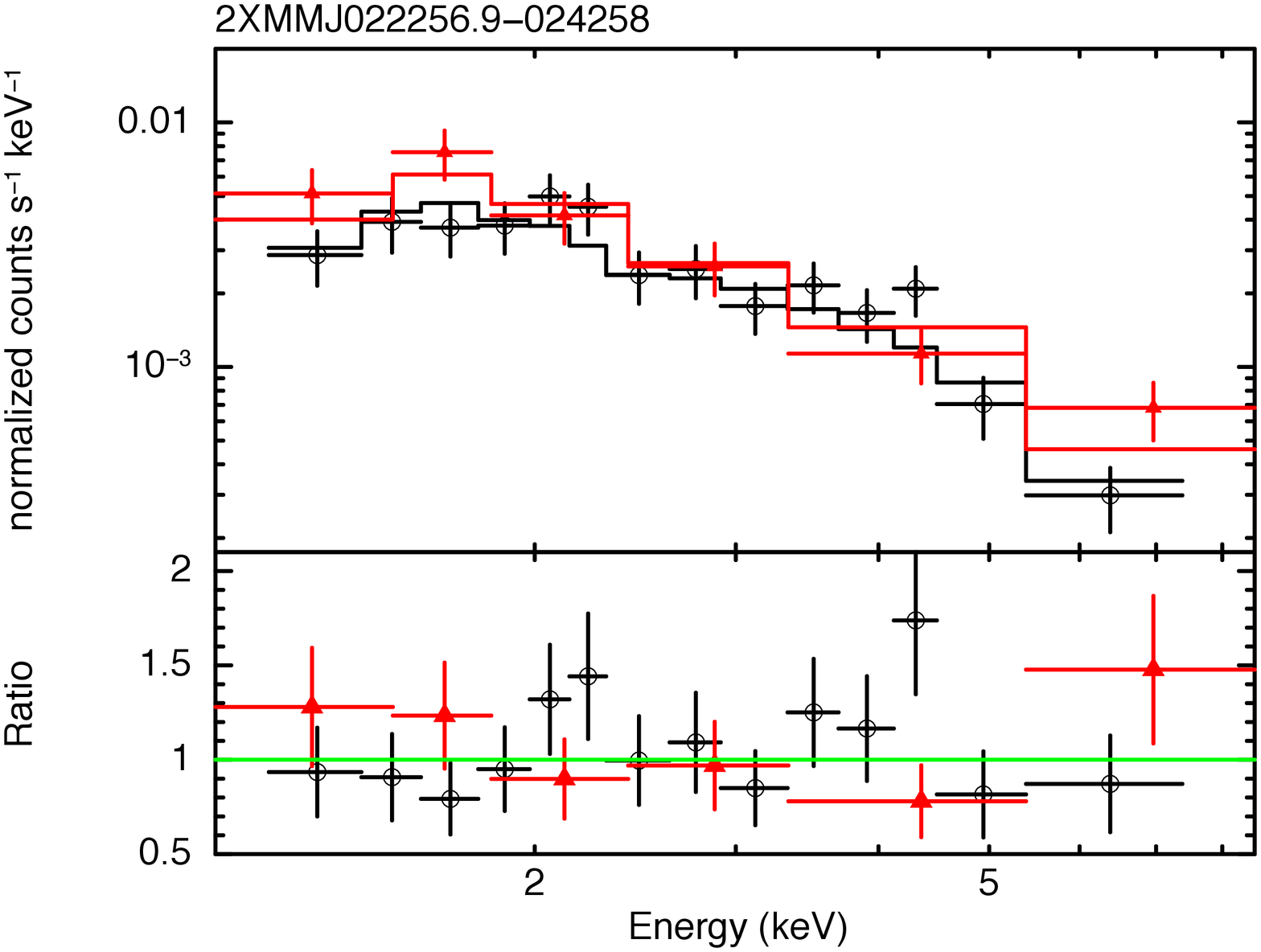}}   
\subfigure{ 
  \includegraphics[width=6cm,angle=0]{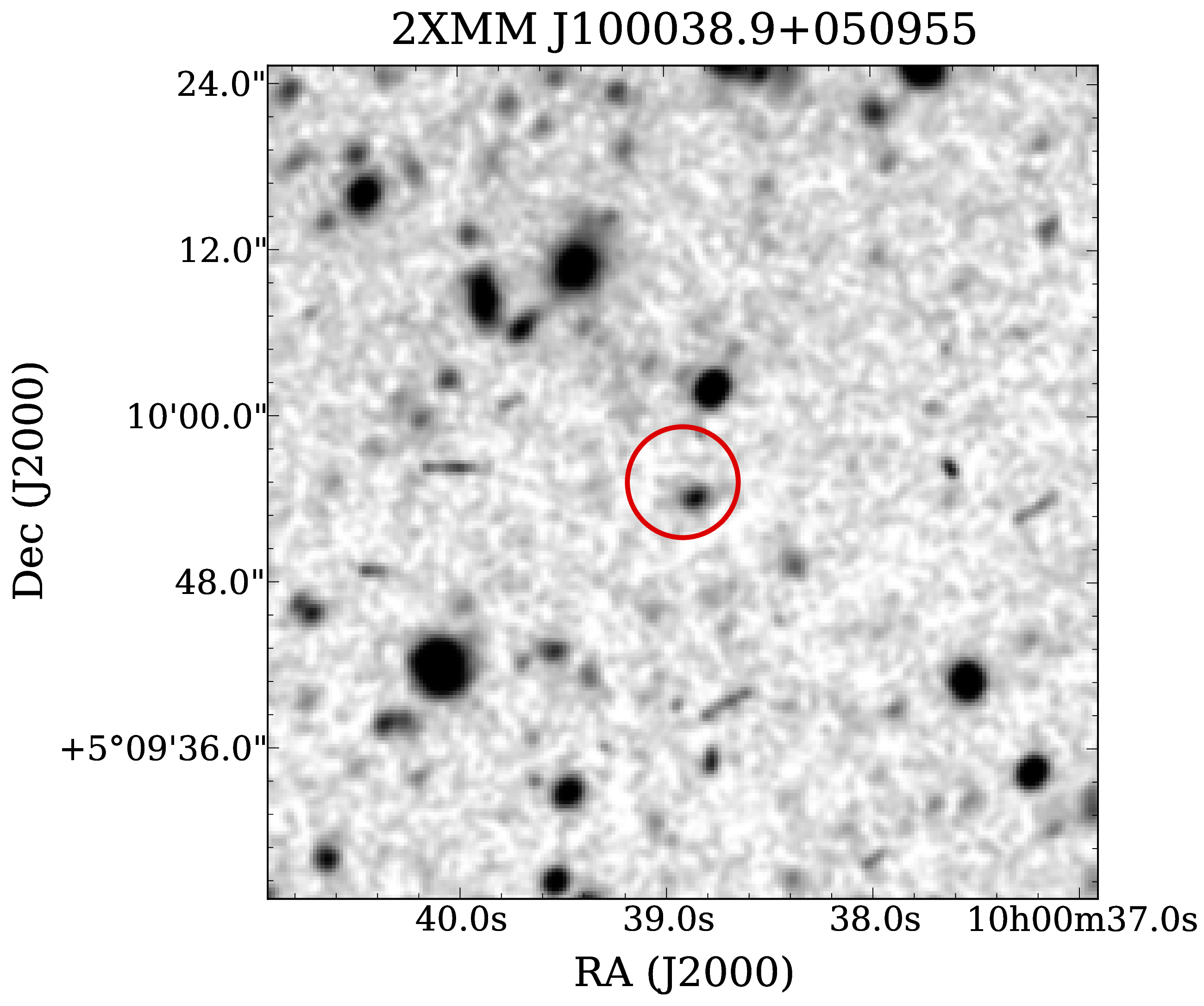}
  \includegraphics[width=6cm,angle=0]{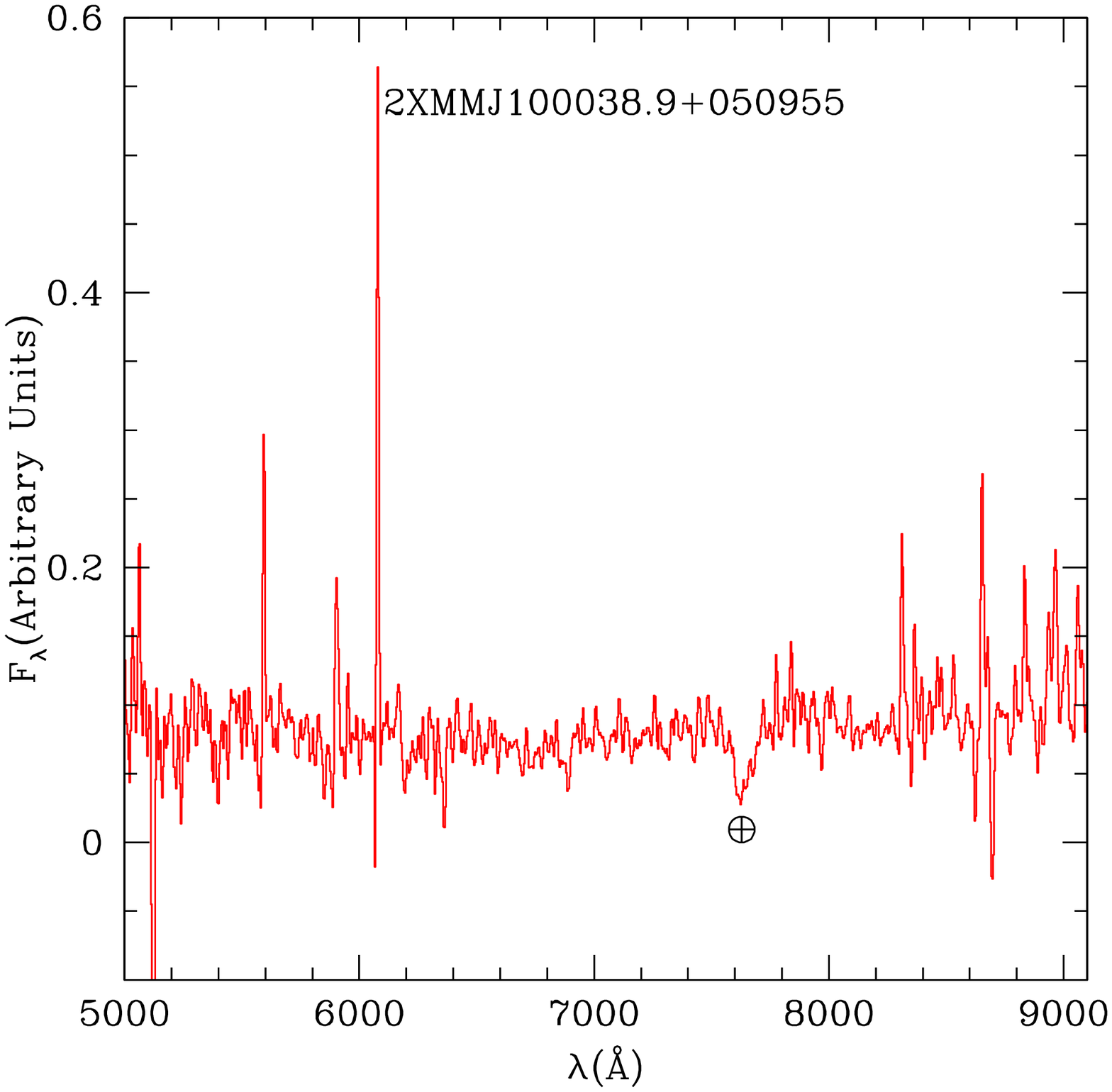}
  \includegraphics[width=6.5cm,angle=0]{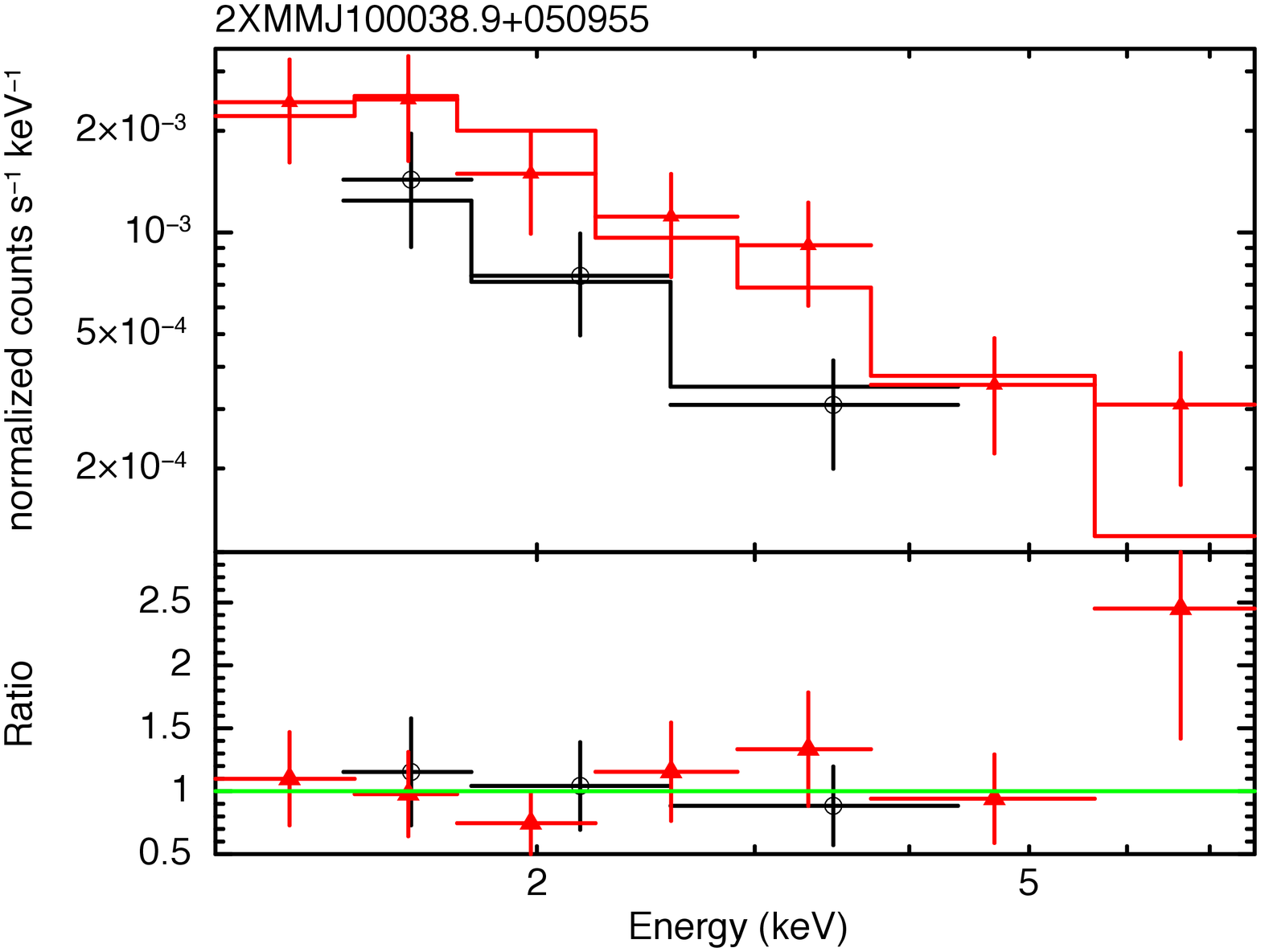}}    
\subfigure{ 
  \includegraphics[width=6cm,angle=0]{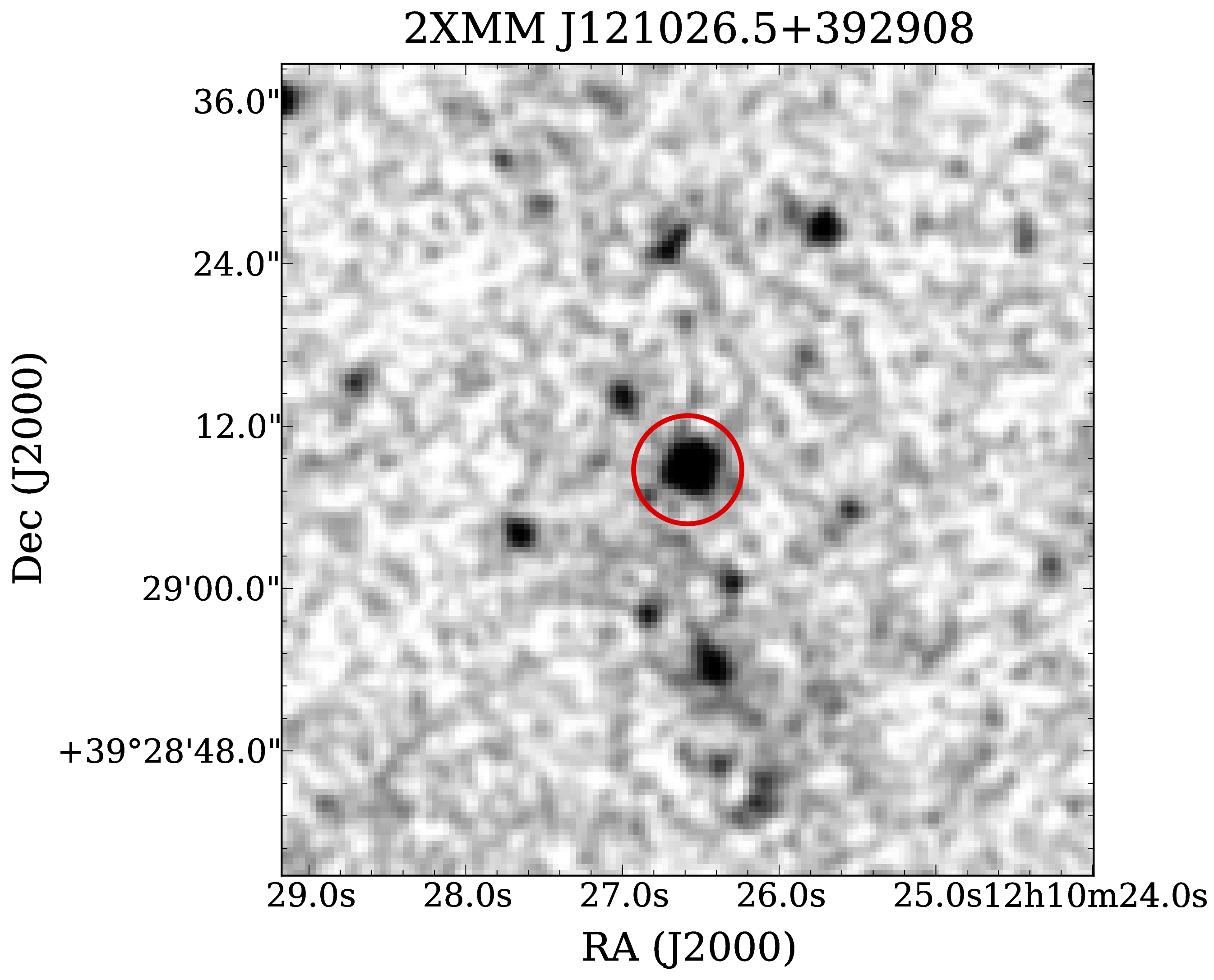}
  \includegraphics[width=6cm,angle=0]{rusco.ps}
  \includegraphics[width=6.5cm,angle=0]{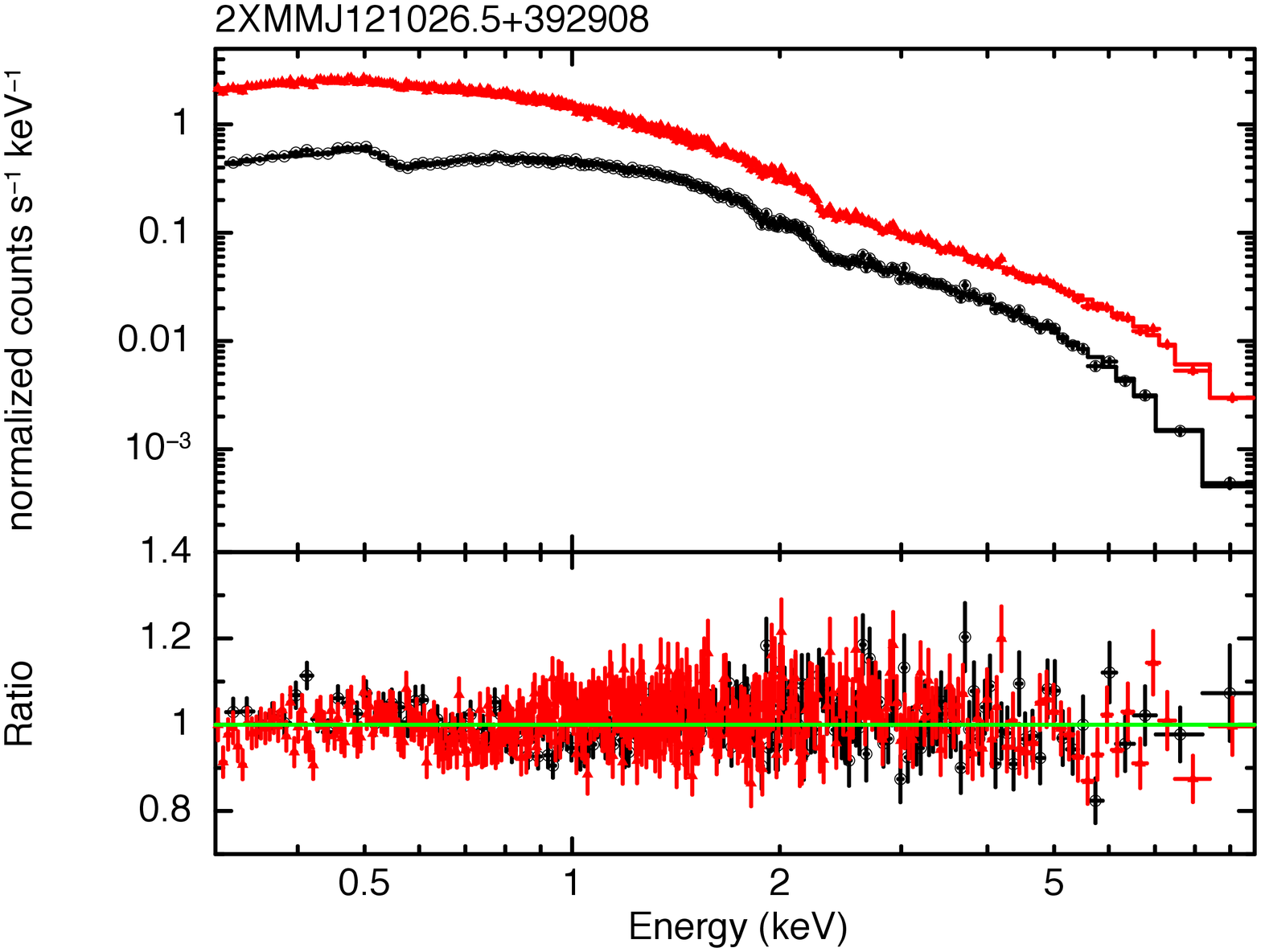}}  
\caption{
{\it Left panels}: Optical finding chart (1x1 arcmin) for the 7 EXO50 objects reported in this paper produced using material from dedicated TNG observing runs (R filter; sources: 2XMM022256.9-024258, 2XMM100038.9+050955, 2XMM123204.9+215254, 2XMM135055.7+642857) or from SDSS (r filter; sources: 2XMM121026.5+392908, 2XMM121134.2+390054, 2XMM143623.8+631726). 
The circle is 4 arcsec radius around the X-ray position.
{\it Middle panels}: The optical spectra of the EXO50 objects observed at the GTC. 
The solid black line in the spectroscopically identified objects represents the underline galaxy component used to reproduce the optical spectrum, except for 2XMMJ121134.2+390054, for which we show the galaxy component (blue line), the AGN component (continuum plus broad emission lines, red line)  and the combined component (galaxy plus AGN; black line). The spectral lines used to classify the objects are also marked. 2XMMJ121026.5+392908 is a well known object 
and its redshift and classification has been taken from the literature. 
 {\it Right panels}: X-ray data (observer frame) and residuals for the EXO50 objects discussed here. All the objects
are well fitted with an absorbed power-law model. Red filled triangles: pn data. Black open circles: MOS data.
}
\end{figure*}
\addtocounter{figure}{-1}
\begin{figure*}
\label{fig}  
\centering
\subfigure{ 
  \includegraphics[width=6cm,angle=0]{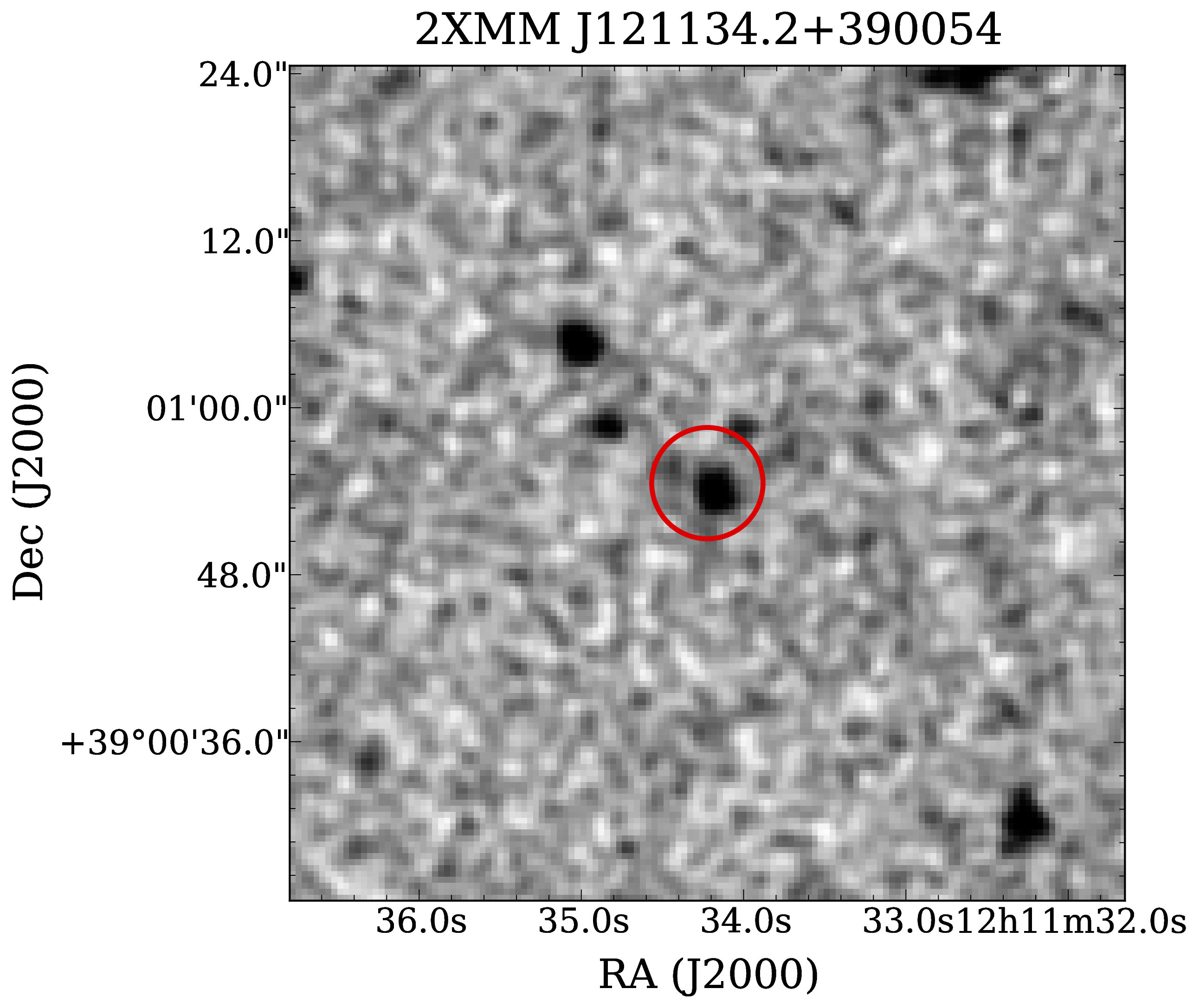}
  \includegraphics[width=6cm,angle=0]{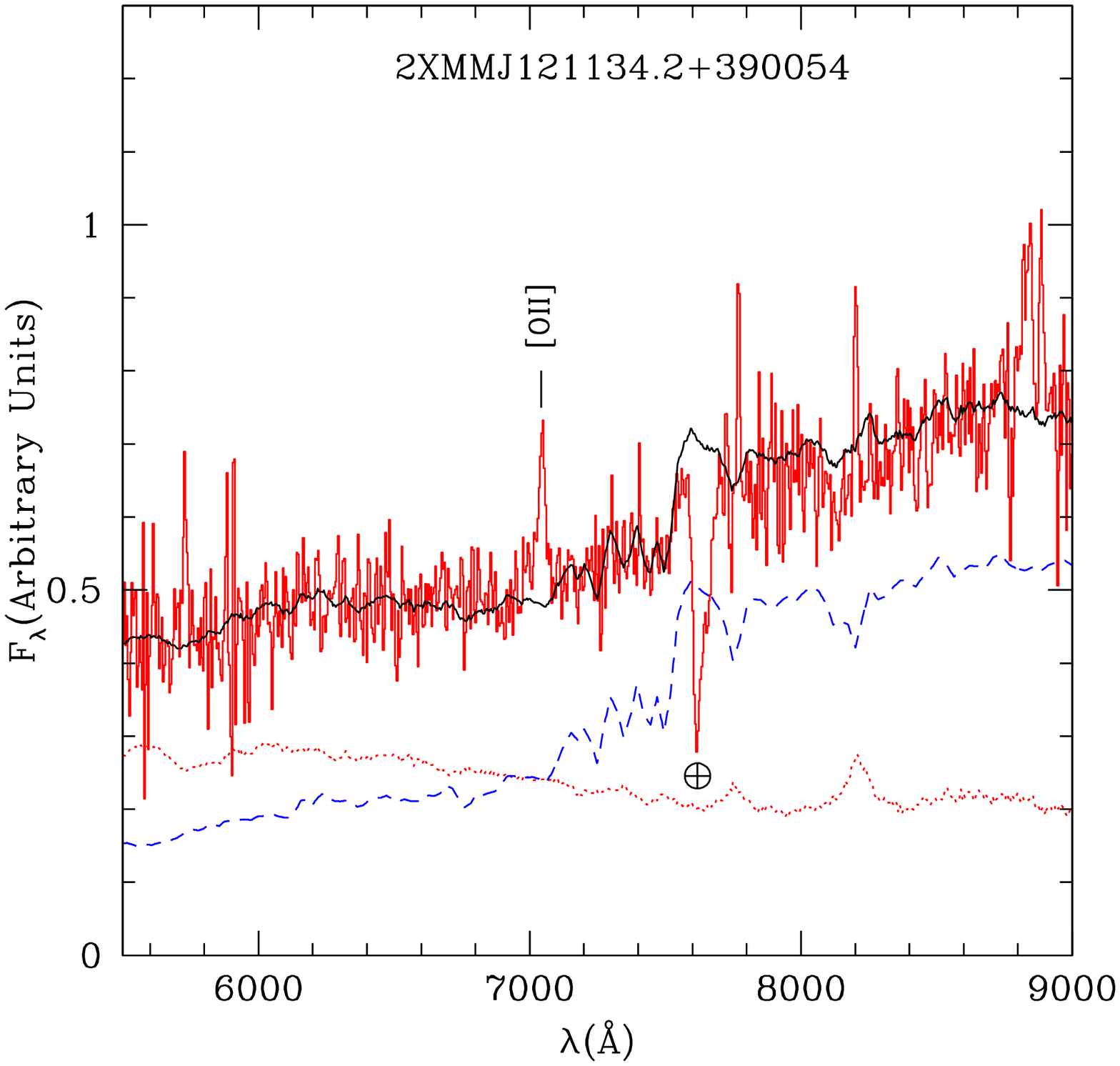}
  \includegraphics[width=6.5cm,angle=0]{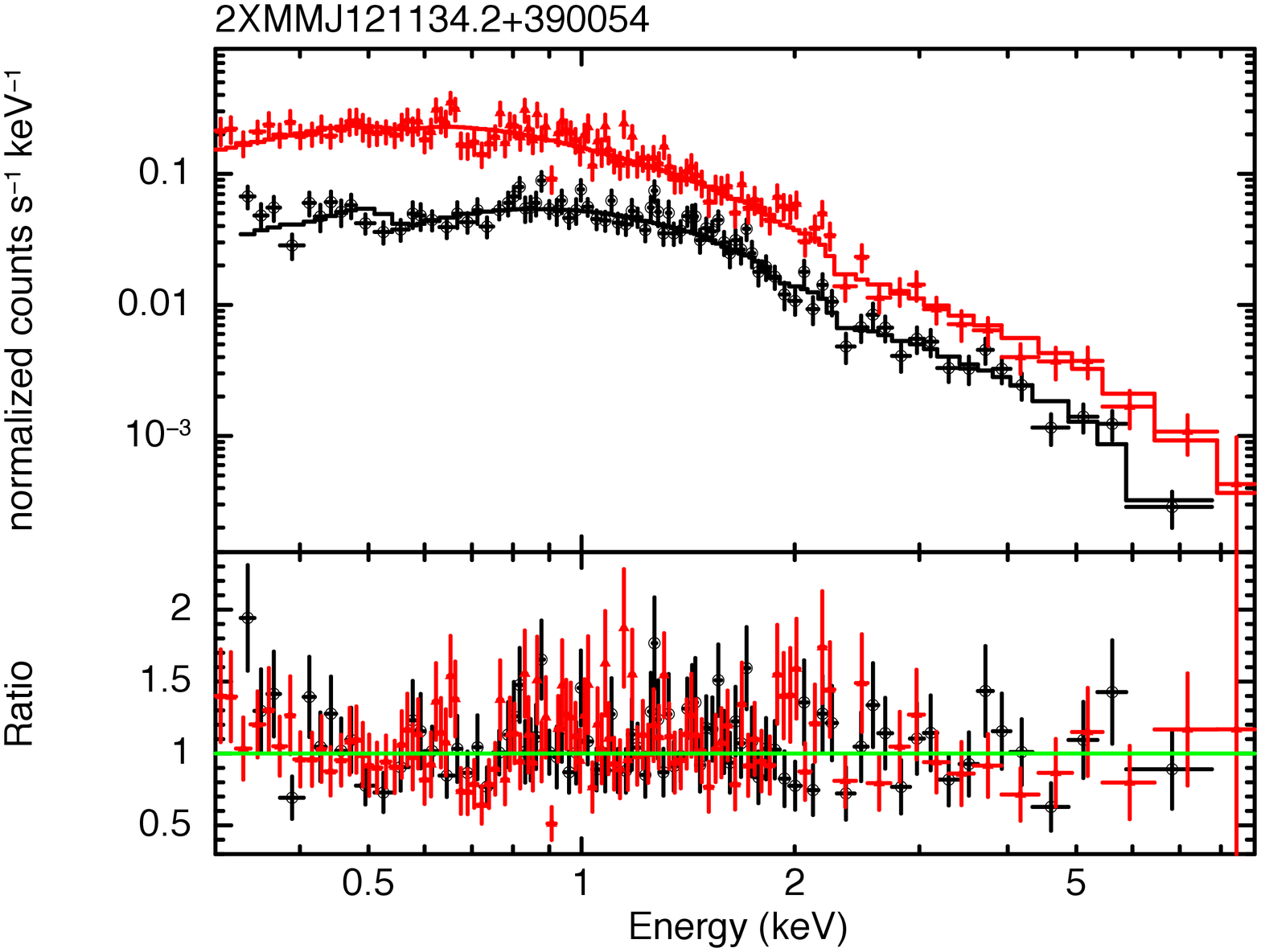}}  
\subfigure{ 
  \includegraphics[width=6cm,angle=0]{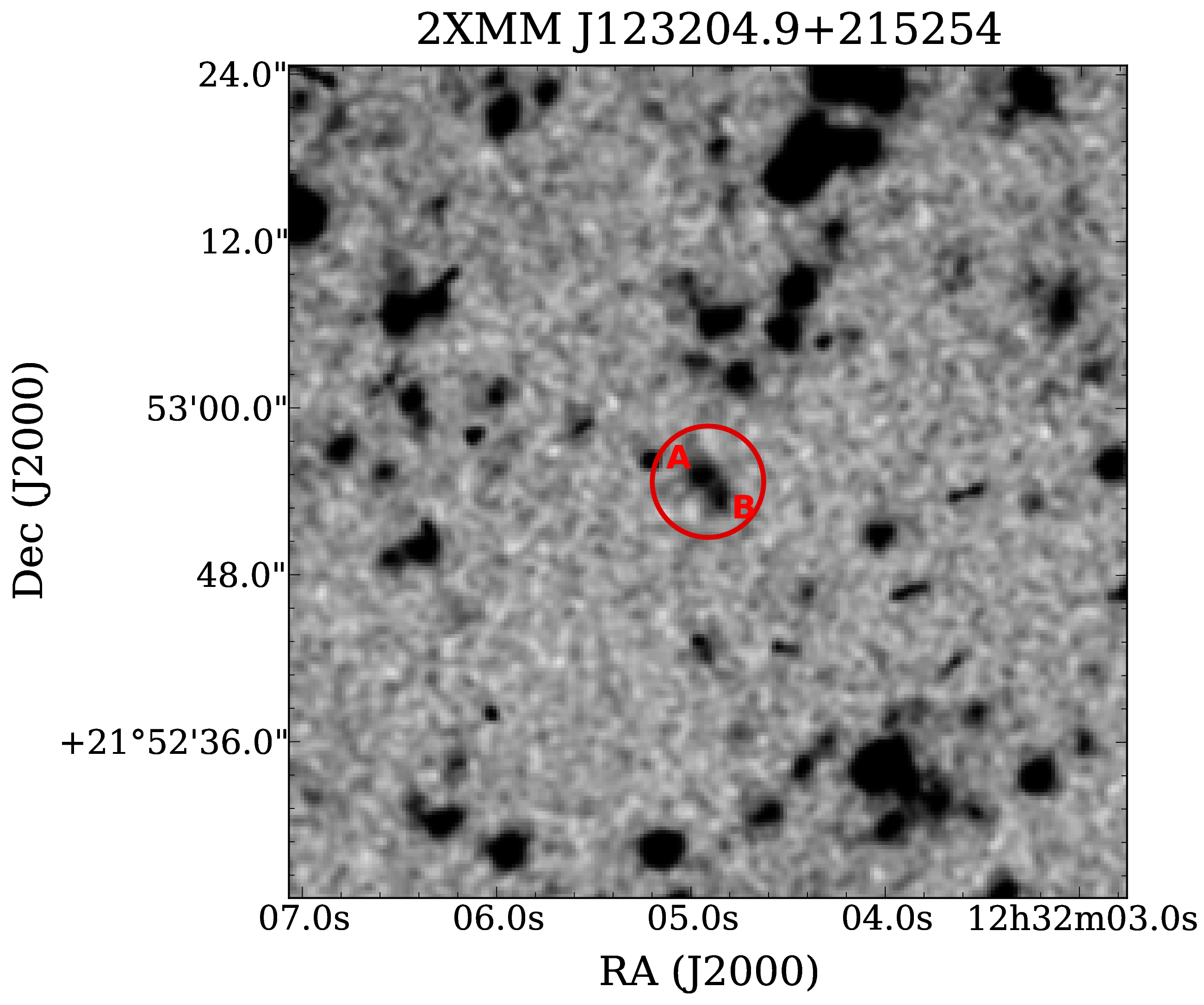}
  \includegraphics[width=6cm,angle=0]{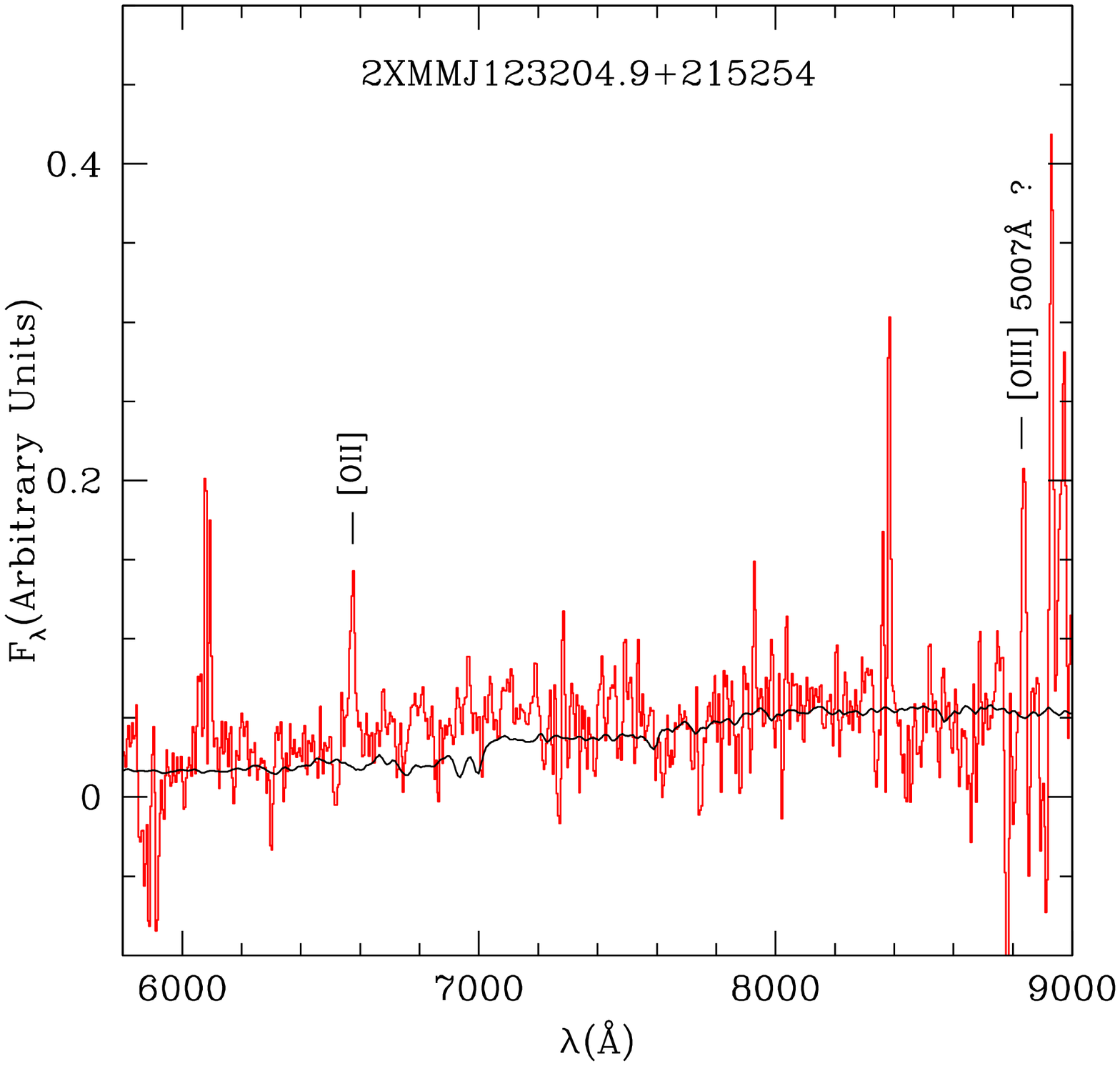}
  \includegraphics[width=6.5cm,angle=0]{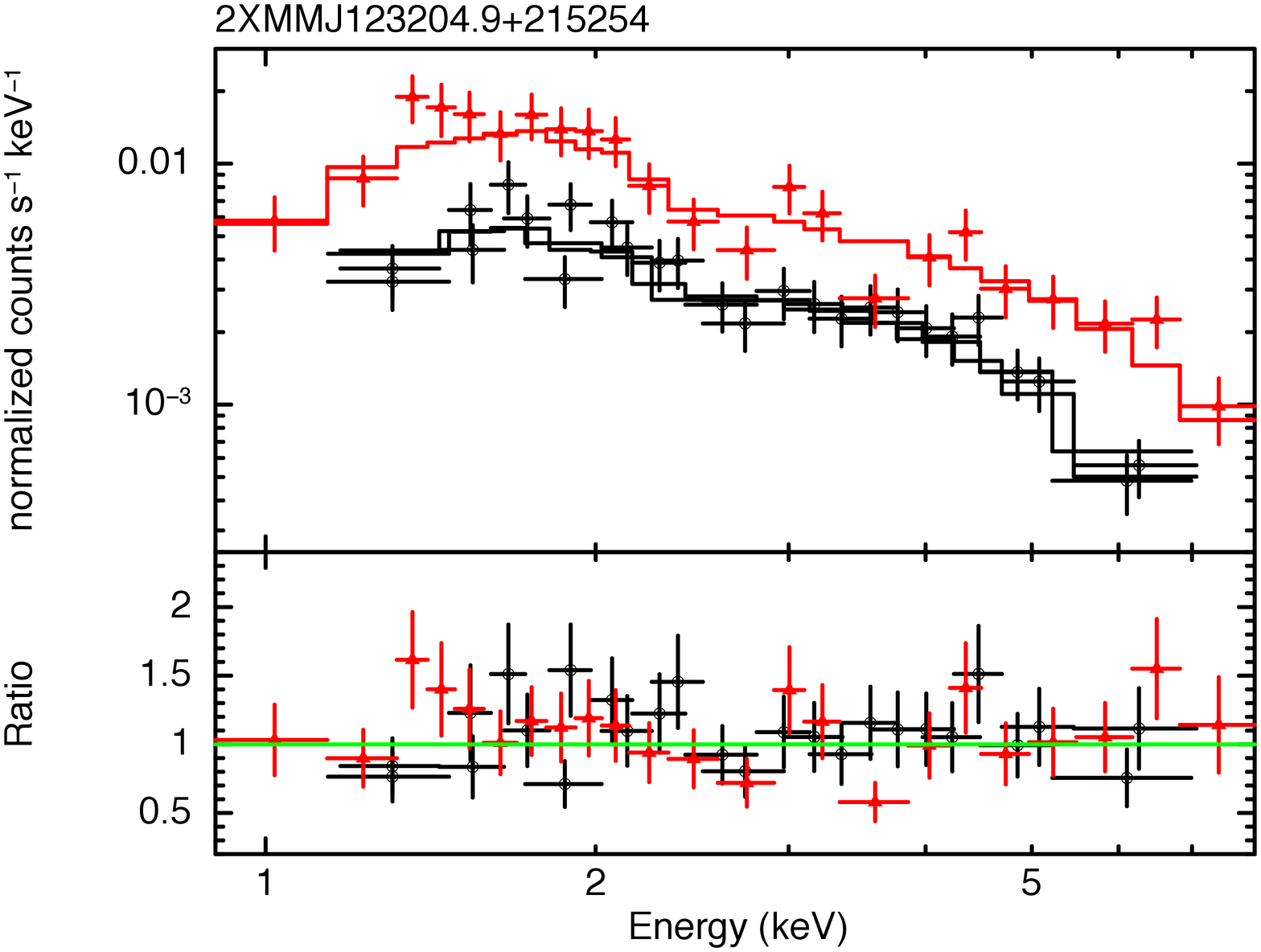}}  
\subfigure{ 
  \includegraphics[width=6cm,angle=0]{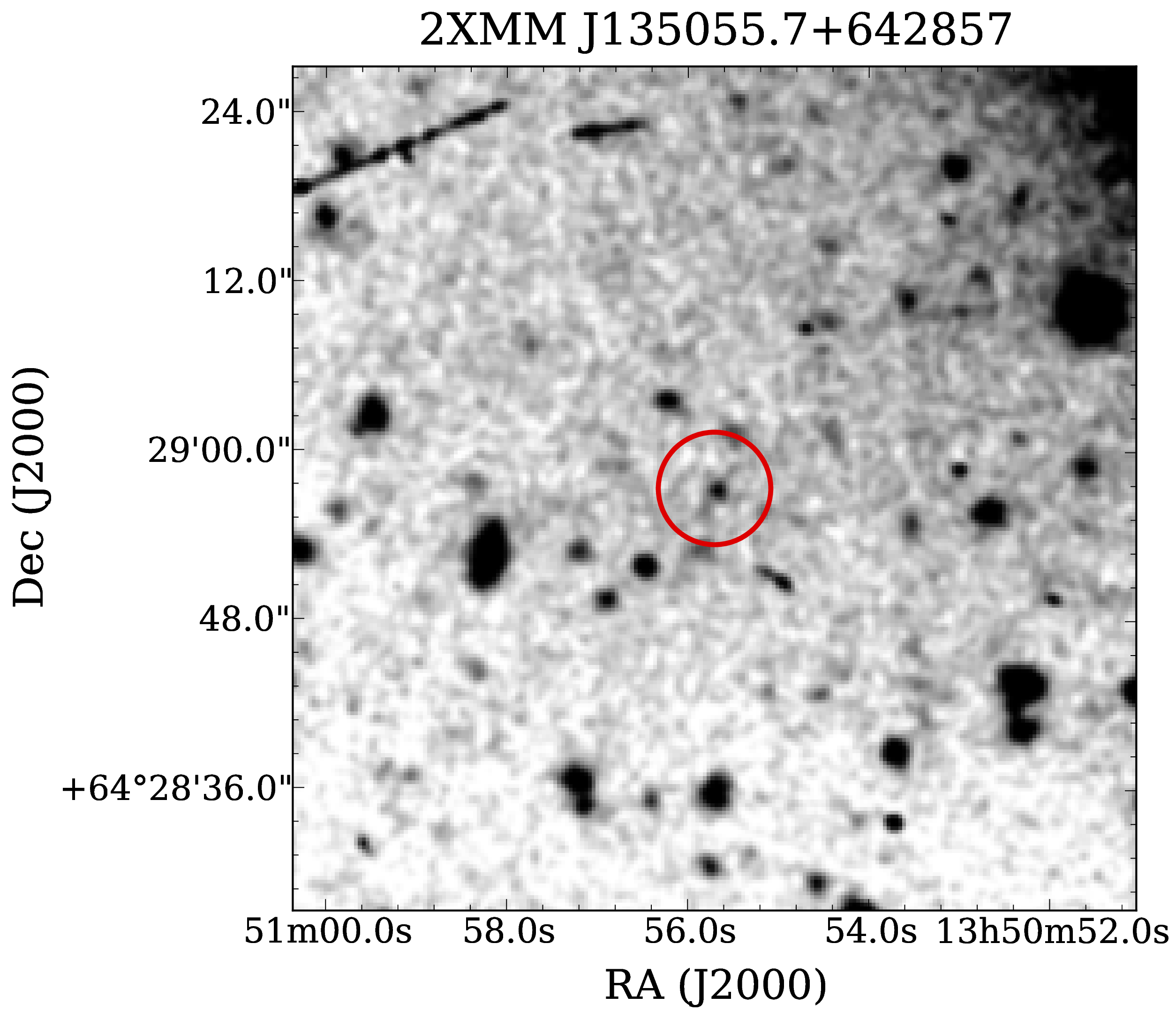}
  \includegraphics[width=6cm,angle=0]{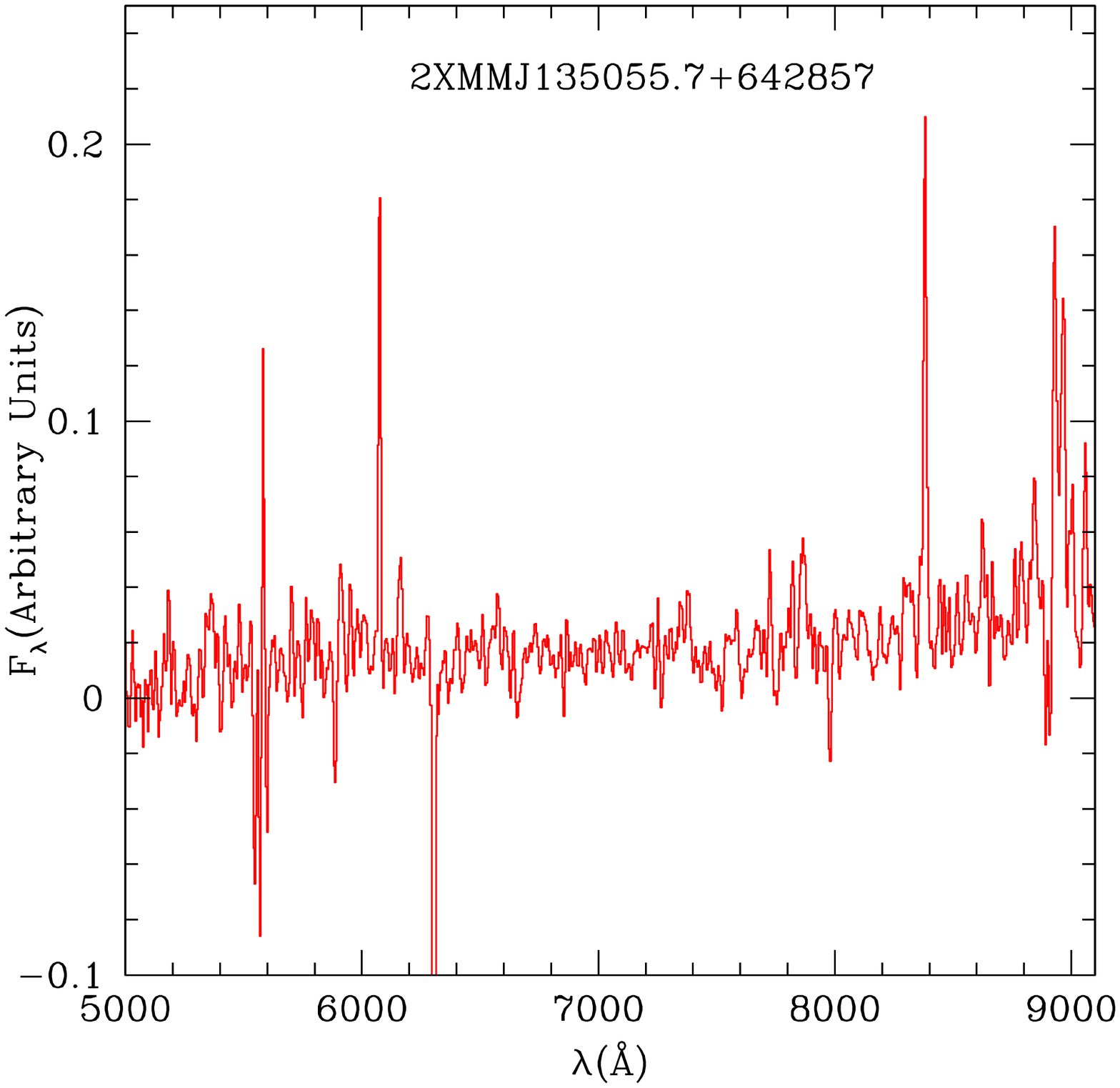}
  \includegraphics[width=6.5cm,angle=0]{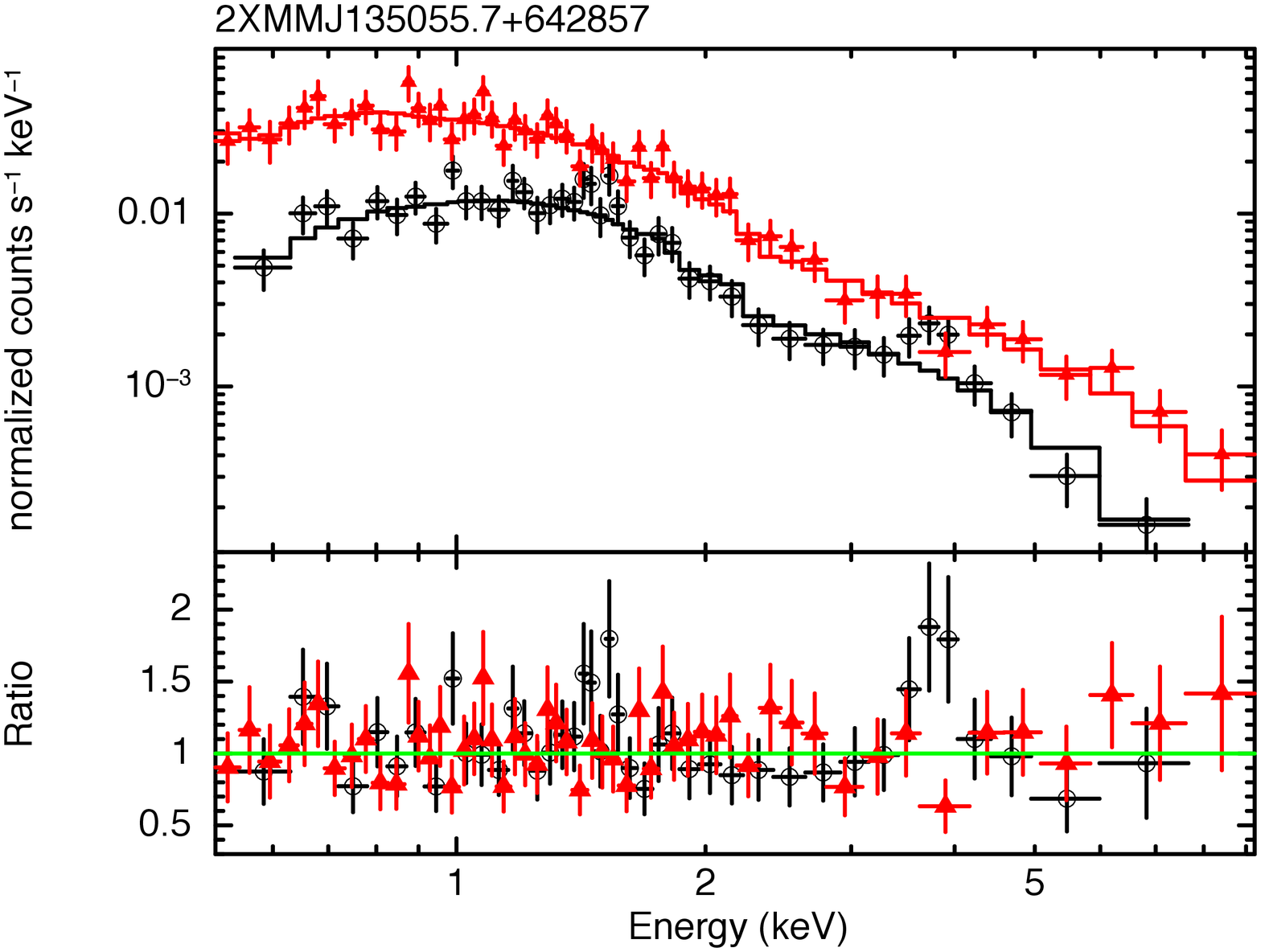}}  
\subfigure{ 
  \includegraphics[width=6cm,angle=0]{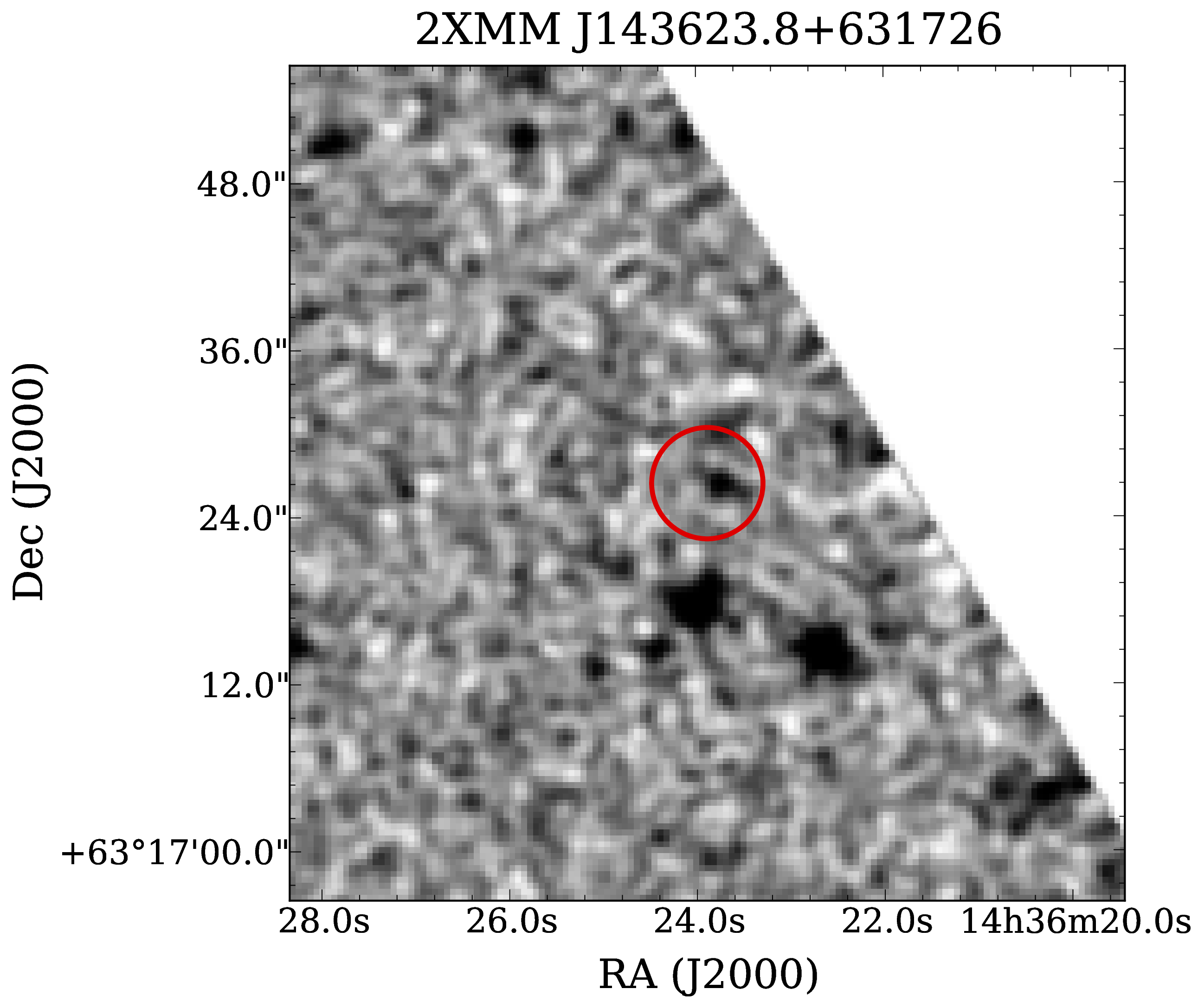}
  \includegraphics[width=6cm,angle=0]{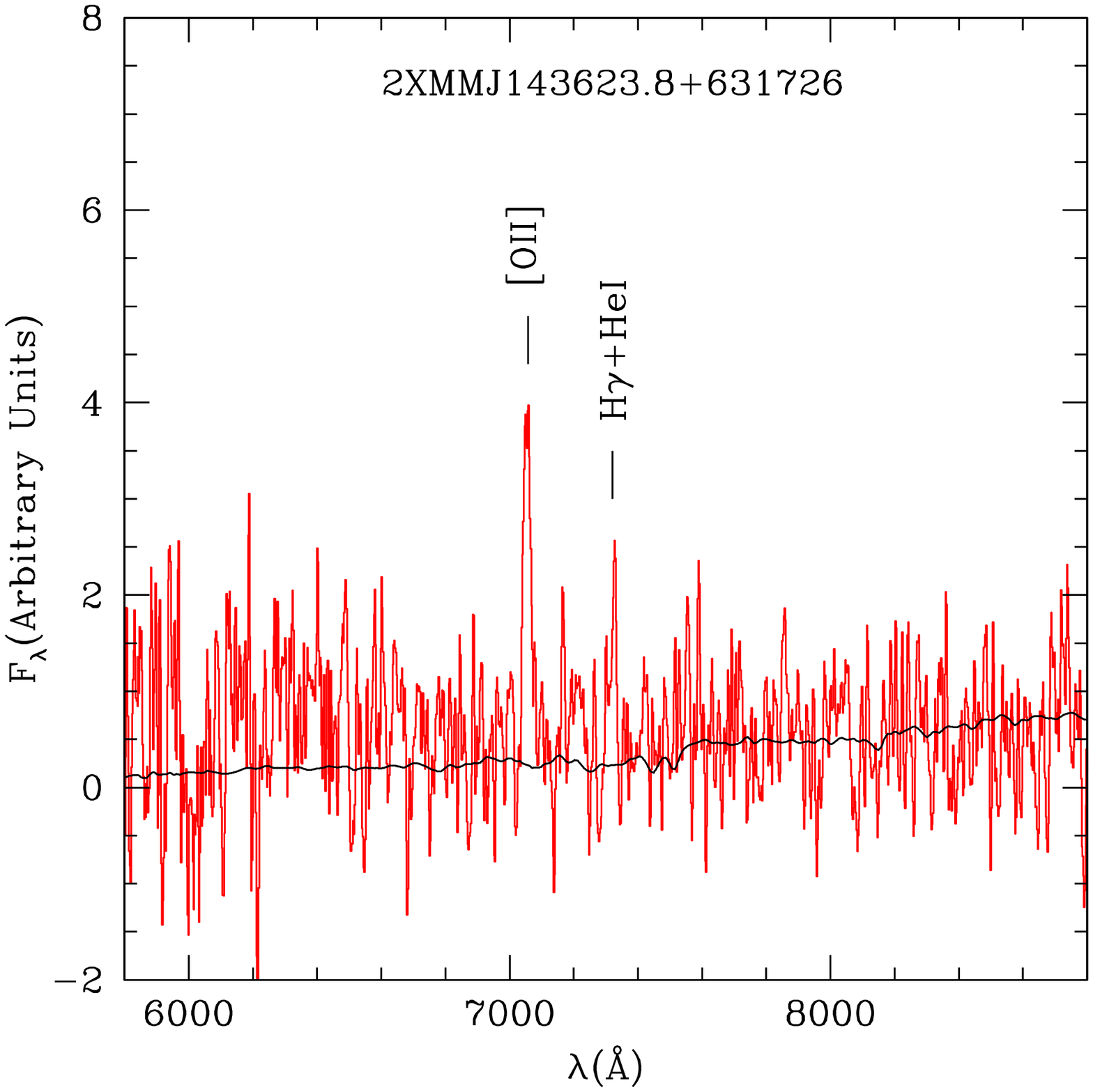}
  \includegraphics[width=6.5cm,angle=0]{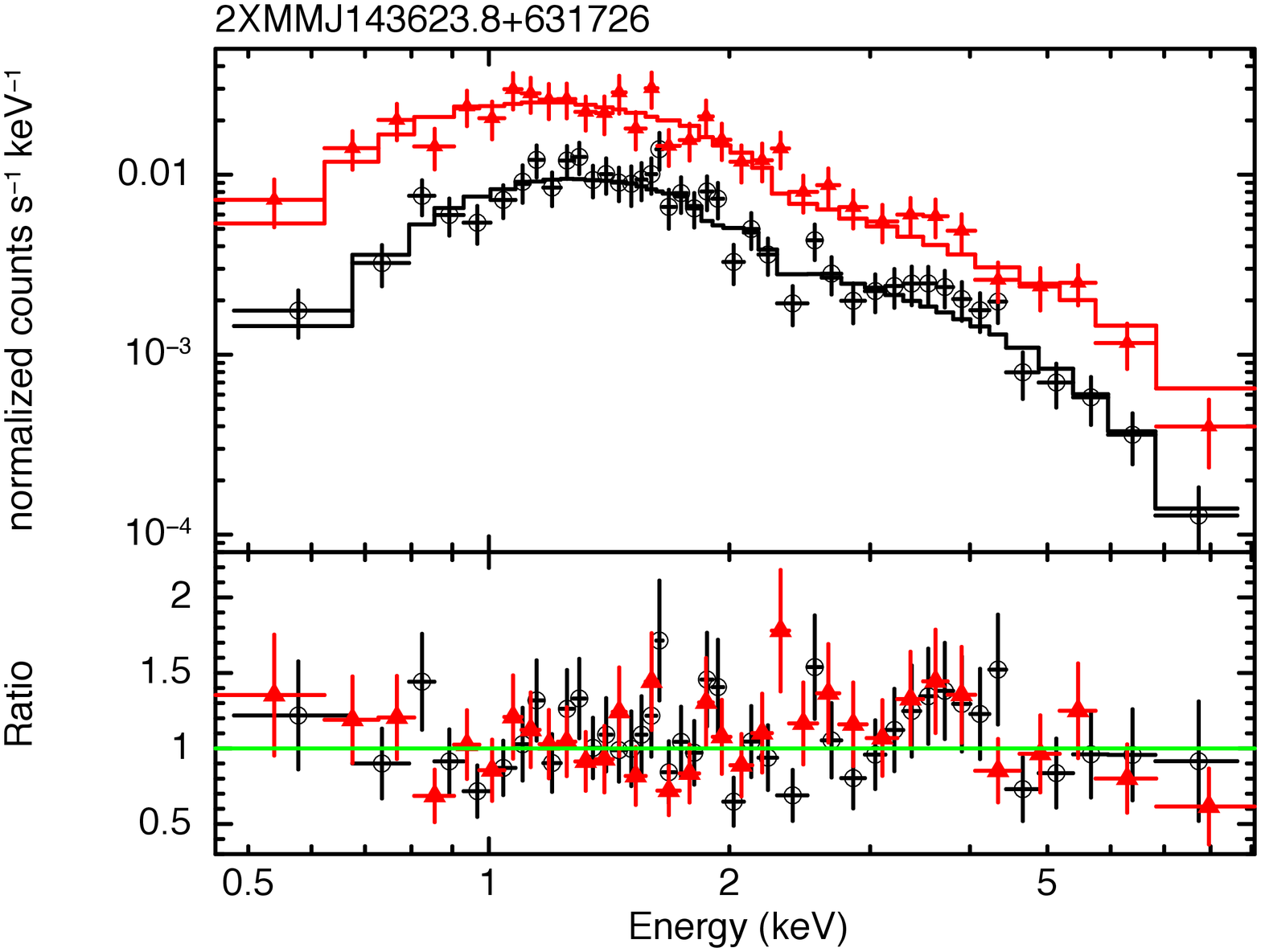}}   
\caption{...continue...}
\end{figure*}

\section{Observations and data Reduction}

\subsection{Optical imaging}

In Figure 2 (left panels) we present the optical finding charts of the 7 sources discussed here (1x1 arcmin) 
produced using data from  dedicated TNG observing runs (R filter) or from the SDSS (r filter). 

TNG raw images were reduced following the standard CCD reduction
process using {\it IRAF} 2.14. All images were de-biased and corrected for
flat-field effects. We combined the different exposures for each source
and the resulting images were flux calibrated using standard stars and
the standard extinction curve for the observatory. Finally, we did an
astrometric calibration of each image; the astrometric accuracy of our 
imaging is typically better than $\sim 0.3$ arcsec. 
The R magnitudes were estimated with {\it IRAF} routines following 
the standard procedure of aperture photometry centered at the source position.

\subsection{Optical spectroscopy and identification}

\begin{table*}
\caption{Optical follow-up spectroscopic observations}             
\label{table:OptObs}      
\centering          
\begin{tabular}{l c l r r c }     
\hline\hline     
2XMMName         & Programme        & Instrument/Grism & Exposure (s)  & Slit Width ($^{\prime\prime}$) & Dates \\ 
(1)              & (2)              & (3)              & (4)           & (5)                            & (6)    \\        
\hline  
J022256.9-024258 & GTC18-10B        & OSIRIS/R500R     & 6$\times$1100 & 1.2        & 2-Sep-2010 (4), 19-Sep-2010 (2) \\
J100038.9+050955 & GTC44-11A        & OSIRIS/R500R     & 4$\times$1800 & 1.0        & 10-Apr-2011 \\
J121134.2+390054 & GTC18-10B        & OSIRIS/R500R     &  2$\times$600 & 1.2        & 27-Jan-2011 \\
J123204.9+215254 & GTC44-11A        & OSIRIS/R500R     & 2$\times$1800 & 1.0        & 10-Apr-2011 (1), 11-Apr-2011 (1) \\
J135055.7+642857 & GTC44-11A        & OSIRIS/R500R     & 3$\times$1800 & 1.0        & 10-Apr-2011 \\
J143623.8+631726 & GTCMULTIPLE2-09A & OSIRIS/R500R     & 7$\times$1500 & 1.2        & 17-Apr-2009 (1), 19-Apr-2009 (3), 25-May-2009 (3)  \\
\hline        
\end{tabular}
Columns are as follows: 
1) Source name;
2) program ID;
3) instrument/grism used
4) number of exposures and exposure times in seconds;      
5) slit width in arcsec; 
6) dates of the observations. In parenthesis we have indicated the number of exposures.
\end{table*}

The optical spectra for 6 of our sources were taken at GTC during 3 different observing runs 
(see Table 2) using the R500R grism on the OSIRIS instrument
\footnote {See http://www.gtc.iac.es/instruments/osiris/osiris.php}.
For the reduction we used the standard IRAF long-slit package following the standard steps.  
The resulting spectra (see Figure 2 middle panels, in arbitrary units) give us spectroscopic information (classification and redshift) for 4 out of the 6 GTC sources. 
For the remaining 2 sources (2XMMJ100038.9+050955 and 2XMMJ135055.7+642857) the spectra are inconclusive and do not 
allow us to derive any relevant information.
The source 2XMMJ121026.5+392908 is a well known object and its redshift and classification have been taken from the literature.
In total we have spectroscopically identified 
(using the GTC data as well as data from the literature) 5 EXO50 objects; two sources 
remain still spectroscopically unidentified, although the data at other wavelengths discussed in section 4 and 5 strongly suggest an obscured QSO nature. 

\subsection{WISE All-Sky Survey data}

The Wide-field Infrared Survey Explorer (WISE; \citealt{wright2010}) has recently carried out an all sky survey in the medium infrared regime, detecting hundreds of millions of objects. The estimated $5\sigma$ point sources sensitivities (in unconfused regions) in the four observed channels (W1=3.4 $\mu$m, W2=4.6 $\mu$m, W3=12 $\mu$m and W4=22 $\mu$m) are better than 0.08, 0.11, 1 and 6 mJy, while the angular resolution (FWHM) are 6.1, 6.4, 6.5, 12 arcsec respectively; the astrometric precision for high signal-to-noise ratio (SNR) sources is better than 0.15 arcsec. 
To add the WISE information to our study  we have cross-correlated our EXO50 sample with the WISE All-Sky source catalog
\footnote{We use here the public available All-Sky Data Release that covers $>$99\% of the sky (March 2012 release;
see http://wise2.ipac.caltech.edu/docs/release/allsky/)}
using a positional tolerance, from the X-ray source position, equal to 4 arcsec . All our sources have been detected by WISE (see below) and {\it a posteriori} the offset between the WISE and the optical position of our sources is below 2 arcsec. 
The results of this cross-correlation are reported in Table 3.
We found a single WISE counterpart for all the 7 sources discussed here. The detections reported in Table 3 have a significance greater than 7$\sigma$ in all bands, with the exception of 2XMMJ100038.9+050955, 2XMMJ123204.9+215254 and 2XMMJ143623.8+631726, detected in the W4 band at $6.1\sigma$, $3.6\sigma$ and $3.9\sigma$ respectively. Two sources, 2XMMJ121026.5+392908 and 2XMMJ121134.2+390054, are not detected in the W3 and W4 bands. 

Using a simulated sample with random positions of $\sim 1600$ object we estimated that the probability to have a random WISE source inside a circle of 2 arcsec radius is $\sim 1\%$ (i.e. 0.07 WISE sources expected by chance in the 7 error circles investigated here), 
implying that all the detected WISE counterparts are very likely associated with the high \fxo sources.

Flux densities at 3.4 $\mu$m, 4.6 $\mu$m, 12 $\mu$m and 22 $\mu$m have been computed from the magnitudes reported in Table 3 
by assuming the magnitude zero points of the Vega system corresponding to a power-law spectrum 
($f_{\nu} \propto \nu^{-\alpha}$) with ${\alpha}=1$ (see \citealt{wright2010}). 
The differences in the computed flux densities expected using flux correction factors that correspond to a 
${\Delta \alpha}=\pm 1$ are lower than 0.8 per cent, 0.6 per cent, 6 per cent, and 0.7 per cent in the W1, W2, W3 and W4 band, respectively (\citealt{wright2010}).

\begin{table*} 
\begin{center} 
\caption{Infrared (WISE) data for the EXO50 sources discussed here.
}
\begin{tabular}{rrrrrrrrrr}
\hline
n & 2XMMName         & \fxo  & ID(z)          & W1              & W2              & W3                &  W4            & Log($\nu L_{\nu}$)  \\        
  &                  &      &                &                 &                 &                   &                &  $(12.3\mu\mbox{m})$  \\        
 (1) & (2)           & (3)  & (4)            & (5)             & (6)             & (7)               & (8)            & (9)           \\   
\hline
\hline
1 & J022256.9-024258 &  72  & AGN2(z=1.004)  & $16.02\pm 0.06$ & $14.89\pm 0.07$ &  $11.08\pm 0.09$  & $8.07\pm 0.15$ & 45.61      \\
2 & J100038.9+050955 & 138  & -              & $16.22\pm 0.08$ & $14.29\pm 0.05$ &  $10.33\pm 0.06$  & $7.97\pm 0.18$ & -           \\            
3 & J121026.5+392908 &  50  & BLLAC(z=0.617) & $14.86\pm 0.04$ & $14.59\pm 0.06$ &  $>12.34$         & $>8.66$        & -  \\
4 & J121134.2+390054 &  50  & BLLAC(z=0.890) & $15.21\pm 0.04$ & $15.03\pm 0.08$ &  $>12.50$         & $>9.45$        & -  \\
5 & J123204.9+215254 & 1118 & AGN2(z=0.763)  & $15.35\pm 0.05$ & $13.84\pm 0.04$ &  $10.19\pm 0.05$  & $8.48\pm 0.30$ & 45.10      \\
6 & J135055.7+642857 & 458  & -              & $14.90\pm 0.03$ & $13.85\pm 0.03$ &  $10.46\pm 0.04$  & $8.47\pm 0.15$ & -           \\    
7 & J143623.8+631726 &  55  & AGN2(z=0.893)  & $15.09\pm 0.03$ & $14.05\pm 0.04$ &  $11.36\pm 0.09$  & $8.91\pm 0.28$ & 45.10     \\
\hline
\end{tabular}
\end{center}
Columns are as follows: 
1) Number used to mark the object in the plots shown in Section 5;
2) source name;
3) X-ray to optical flux ratio; 
4) optical spectroscopic classification and redshift;
5) W1 magnitude (3.4 $\mu$m) and 1$\sigma$ error;
6) W2 magnitude (4.6 $\mu$m) and 1$\sigma$ error;
7) W3 magnitude (12 $\mu$m) and 1$\sigma$ error. 
The magnitude lower limits correspond to the 95\% confidence level as reported in the WISE catalog;
8) W4 magnitude (22 $\mu$m) and 1$\sigma$ error.
The magnitude lower limits correspond to the 95\% confidence level as reported in the WISE catalog;
9) Log of the rest frame 12.3 $\mu$m luminosities ($\nu_{12.3 \mu m} \times L_{12.3 \mu m}$ in \es) computed as described in Section 3.3.
\end{table*}

\subsection{XMM-Newton spectroscopy}

\begin{table*}
 \caption{EPIC {\xmm} observation details.}             
 \label{tab:xmmlog}      
 \begin{center}
 {
  \begin{tabular}{l@{\extracolsep{0.1cm}} l@{\extracolsep{0.1cm}} c@{\extracolsep{0.1cm}} c@{\extracolsep{0.1cm}} c@{\extracolsep{0.2cm}} c@{\extracolsep{0.2cm}} r}
  \hline\hline       
    \multicolumn{1}{c}{Name} & \multicolumn{1}{c}{{\sc OBSID}} & $N_H$ & Instr.    & Filter & Net Exp. Time  & Net Counts \\
     \multicolumn{1}{c}{(1)} & \multicolumn{1}{c}{(2)} & (3) & (4) & (5) & (6) & (7) 
    \vspace{0.1cm} \\
    \hline
    \vspace{-0.2cm} \\
     J022256.9$-$024258      & $0037981601$                    & 2.34  & MOS1\&2   & thin   & 26.1           &   265 \\
                             &  	                           &       & pn        & thin   &  7.9           &   127  \\ 
     \hline   
     J100038.9+050955        & $0204791101$                    & 2.41  & MOS1      & med    & 16.1           &    27  \\
                             &  	                           &       & pn        & med    & 12.5           &    63  \\ 
     \hline   
     J121026.5+392908        & $0112830501,$                   & 2.00   & MOS1\&2  & med    & 161.5          &    119876  \\
                             & $0112830201$	                   &        & pn       & med    &  68.9          &    175724  \\ 
     \hline   
     J121134.2+390054        & $0112190201$                    & 1.84  & MOS1\&2   & med    & 27.8           &    2311  \\
                             &  	                           &       & pn        & med    &  8.9           &    2488  \\ 
     \hline     
     J123204.9+215254        & $0112650301$                    & 1.80  & MOS1      & thin   & 19.3           &    248  \\
                             &  	                           &       & MOS2      & med    & 19.5           &    250  \\ 
                             &  	                           &       & pn        & thin   & 13.1           &    447  \\ 
     \hline   
     J135055.7+642857        & $0147540101$                    & 1.70  & MOS1\&2   & med    & 44.6           &    873  \\
                             &  	                           &       & pn        & med    & 18.9           &   1064  \\ 
     \hline   
     J143623.8+631726        & $0204400301$                    & 1.37  & MOS1\&2   & med    & 46.7           &    862  \\
                             &  	                           &       & pn        & med    & 12.7           &    648  \\ 
     \hline   
     \vspace{-0.2cm}     
  \end{tabular}
 }
 \end{center}       
 {
 \footnotesize Col. (1): 2XMM source name;
 \footnotesize Col. (2): {\sc OBSID} of the \xmm\ observation;
 \footnotesize Col. (3): Galactic absorbing column density along the line of sight in units of $10^{20}$ cm$^{-2}$;
 \footnotesize Col. (4): EPIC instrument;
 \footnotesize Col. (5): EPIC filter;
 \footnotesize Cols (6): Exposure time after removing high-background intervals, in units of ksec;
 \footnotesize Col. (7): Net counts in the energy range $0.3 - 10\,$keV.}
\end{table*}

In Table 4 we report details for the XMM-{\it Newton} data used for the X-ray spectral analysis of each source discussed here. The XMM-Newton data were cleaned and processed with the XMM-Newton Science Analysis Software (SAS) and were analyzed with standard software packages (FTOOLS; XSPEC, \citealt{arnaud1996}). 
Event files produced from the pipeline were filtered from high-background time intervals and only events corresponding to patterns 0-12 for MOS and 0-4 for pn were used. All spectra were accumulated from a circular extraction region with a radius of 20-30 arcsec, depending on the source off-axis distance.
Background counts were accumulated in nearby circular source free regions, using an area usually about a factor of 4 larger than the one used to extract the source counts. The X-ray spectra usually cover the 0.3-10 keV energy range; the total (MOS1+MOS2+pn) net (background subtracted) counts range from $\sim 10^2$ to $\sim 3\times 10^5$ counts. 
The ancillary response matrix and the detector response matrix were created by the XMM-SAS tasks {\it arfgen} and {\it rmfgen} at each source position in the EPIC detectors. 
To improve the statistics, the MOS1 and MOS2 spectra obtained by using the same filter were combined a posteriori by using the FTOOLS task {\it mathpha}; in this case ancillary and detector response matrices for the MOS1 and MOS2 detectors for each source were combined by using the tasks {\it addrmf} and {\it addarf}. For all the sources but 2XMMJ100038.9+050955 we grouped the spectra in bins containing more than 20 (source+background) counts and used the $\chi^2$ minimization technique; in the case of 2XMMJ100038.9+050955 we grouped the spectra in bins containing 10 (source+background) counts and use the Cash statistics. In the case of 2XMMJ121026.5+392908 two XMM observations were used; pn (MOS) data from the two independent data set were combined together and the ancillary and detector response matrices were created using the same procedures quoted above. We fitted pn and MOS spectra simultaneously in the 0.3-10 keV band, tying together all pn and MOS parameters except for a relative normalization, which accounts for the differences between pn and MOS flux calibrations (see \citealt{mateos2009}). In the following, derived fluxes and luminosities refer to the MOS instrument. 

For the spectral modelling we considered a simple absorbed power-law model that takes into account both the Galactic hydrogen column density along the line of sight (from \citealt{dickey1990}) and a possible intrinsic absorption at the source redshift (abundances relative to the Solar one as reported
in \citealt{wilms2000}). In the X-ray spectral modelling we made use, when available, of the redshifts obtained from the optical spectroscopy. All the X-ray spectra are well fitted by this simple model; the results are reported in Table 1 along with the corresponding flux (corrected for Galactic absorption) and intrinsic luminosity (i.e. corrected for both 
Galactic and intrinsic absorption) in the standard 2-10 keV energy band. The X-ray spectra are shown in Figure 2 (right panels).
In the case of 2XMMJ100038.9+050955, the spectral quality does not allow us to constrain the power-law photon index and the intrinsic absorption at the same time, so we have fixed the power-law photon index $\Gamma$ to 1.9, a common value for unabsorbed AGN ({\citealt{mateos2010}, \citealt{corral2011}, \citealt{lanzuisi2013}). 

\section{Results}

Before discussing the properties of each single object in this sample we summarise here their main properties. 
From optical spectroscopy, out of the 7 EXO50 sources 2 are classified as BL Lac objects, 3 are classified as Type 2 QSO and 2 remain unidentified.
The 3 sources classified as Type 2 QSO are in the redshift range 
0.7$-$1 and are characterised by an intrinsic X-ray absorbing column density, $N_H$, in the range between 1.5 and 8 $\times 10^{22}$ cm$^{-2}$. For the two unidentified objects the X-ray analysis provides lower limits to the 
intrinsic $N_H$ of $4\times 10^{21}$ cm$^{-2}$ and $10^{21}$ cm$^{-2}$, respectively. There are no Compton Thick AGN ($N_H$ in excess to $\sim 10^{24}$ cm$^{-2}$) amongst the 3 Type 2 QSO, neither is suspected their presence in the 2 still spectroscopically unidentified sources, since their X-ray spectra are at odds with that usually observed in Compton Thick AGN (e.g. the presence of a prominent Iron line at 6.4 keV, rest frame, or a very flat X-ray spectrum). The 2 BL Lac objects are at z=0.62 and z=0.89 and, as it will be discussed in Section 5.7, they are rather extreme in their SED. 

For the three EXO50 sources spectroscopically identified as Type 2 QSOs (see below), 
we report in Table 3 the rest frame 12.3 $\mu$m luminosity obtained by interpolating 
the observed luminosity (i.e. not corrected for reddening) in the W3 (observed frame, 12 $\mu$m) and  W4 (observed frame, 22 $\mu$m) WISE bands and using the measured spectral index 
between 12 $\mu$m and 22 $\mu$m;  the same spectral index has been used to evaluate the K-correction. 
\cite{gandhi2009}, studying a sample of Type 1 and Type 2 Seyferts, showed that the observed luminosity around 12.3 $\mu$m rest frame ($\nu_{12.3 \mu m} \times L_{12.3 \mu m}$)
should represent an accurate proxy for the AGN intrinsic power; 
in powerful AGN, the contribution expected from the host galaxy at these wavelengths is marginal ($< 10\%$; \citealt{ballo2014}). 
Overall the derived infrared luminosities are in the range typical of 
Luminous Infrared Galaxies  (LIRG: $L_{IR} > 10^{11} L_{\odot}$) and 
Ultra Luminous Infrared Galaxies  (ULIRG: $L_{IR} > 10^{12} L_{\odot}$, see e.g. 
\citealt{sanders1996}).
Assuming a bolometric correction factor to the 12.3 $\mu$m luminosity of $\sim 10.8$, appropriate for high luminosity AGN (see \citealt {ballo2014}) the implied bolometric luminosities for the three spectroscopically identified  
Type 2 QSO are in the range $1.5-4.5 \times 10^{46}$ \es.

\begin{itemize}

\item{\underline {2XMMJ022256.9$-$024258; \fxo=72}} 

A single source (R=22.1) is evident in the optical finding chart within the X-ray error circle (see Figure 2).
The optical spectrum of this object shows several narrow (observed FWHM $<1000-1500$ \kms) emission lines that we associate with 
MgII$\lambda2798$, 
[NeV]$\lambda3346,3426$, 
[OII]$\lambda3728$,
[NeIII]$\lambda3869$, 
H$_\epsilon$, 
H$_\delta$, 
H$\gamma$, 
[OIII]$\lambda4364$ 
and HeI$\lambda4471$
at $z=1.004$. 
The permitted emission lines (H$\gamma$ and 
MgII$\lambda2798$) have widths similar to the forbidden ones. The H$\gamma$ has a FWHM of 550-600 km/s that is close to the instrumental resolution. The MgII$\lambda2798$ has a strongly asymmetric profile, which is difficult to analyse with the present data, being 
the line at the border of the sampled wavelength range; the FWHM of this line is likely below 1500 km/s. 
It is worth noting the presence in this object of the  [NeV]$\lambda3426$ line, a reliable signature of nuclear activity since it can not be produced in starburst/starforming galaxies (see e.g. \citealt{gilli2010}, \citealt{mignoli2013}). The X-ray spectrum is well described by an
absorbed power-law model with an $N_H$ of $\sim 7.5 \times 10^{22}$ \cm; the intrinsic, rest-frame, 2-10 keV luminosity is $\sim 1.9 \times 10^{45}$ \es. 
Based on the optical and X-ray spectral properties (optical line widths, intrinsic $N_H$ and 
intrinsic 2-10 keV luminosity in excess to $10^{44}$ \es) we classify this source as a Type2 QSO.
2XMMJ022256.9$-$024258 is detected in all the WISE bands (see Table 3).

\begin{figure}
\begin{center}
\resizebox{0.46\textwidth}{!}{
\rotatebox{0}{
\includegraphics{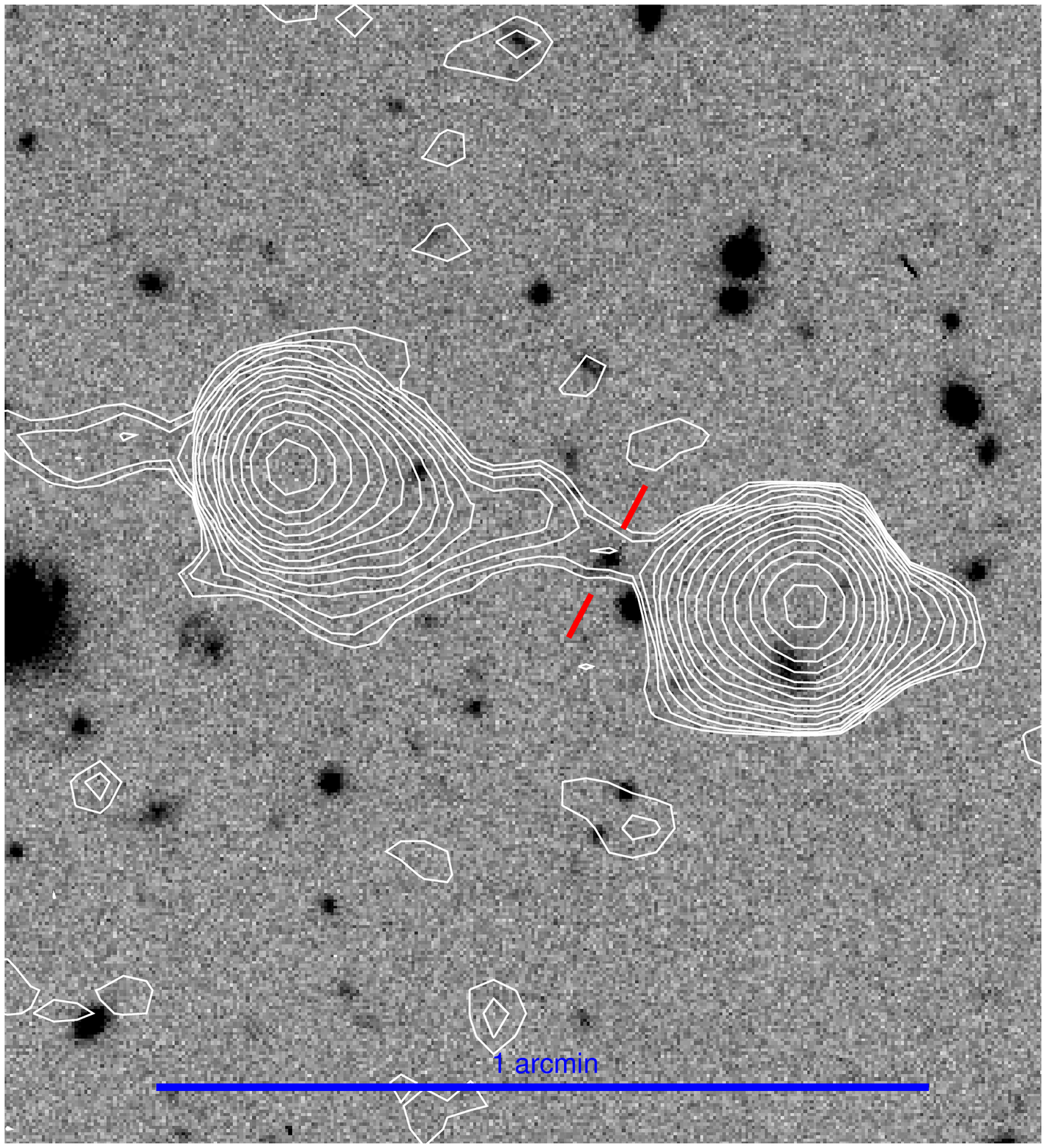}
}}
\caption{Radio (FIRST) contour plots superimposed to a CCD R image (taken at the TNG) of 2XMMJ022256.9$-$024258; the red hatch indicates the object responsible for the X-ray emission. The radio contours start from 0.2 mJy with a multiplicative step of $\sim 1.58$. The large blue bar at the bottom is 1$^{\prime}$ wide: at the redshift of the source 1$^{\prime}$ corresponds to $\sim 0.48$ Mpc. North is upwards, East is on the left.}
\label{fig}
\end{center}
\end{figure}

2XMMJ022256.9$-$024258 also has strong radio emission associated with the radio source 4C-02.11.
The total 1.4 GHz radio flux from the FIRST survey (\citealt{white1997}) is $\sim 340$  mJy corresponding to a
radio power of  $\sim 1.7\times 10^{34}$ erg s$^{-1}$ Hz$^{-1}$ at z=1.004. 
The source is also detected in the lower spatial resolution 
NVSS survey (\citealt{condon1998}). The total NVSS radio flux density (power) is
$429\pm 14$ mJy ($2.18\pm 0.07 \times 10^{34}$ erg s$^{-1}$ Hz$^{-1}$); 
the radio flux density from the NVSS is a more reliable measure of the 
total radio flux, since the VLA-B configuration used for the FIRST survey could 
miss some of the diffuse extended emission. 
For this source we also found in the literature fluxes at 178 MHz
($f_{178 MHz} = 2.6\pm 0.65$ Jy; \citealt{gower1967})
and 4.85 GHz 
($f_{4.85 GHz} = 0.116\pm 0.012$ Jy; \citealt{griffith1995}); 
assuming a power-law model the derived radio spectral index is 
$\sim 0.94$, consistent with the typical values of lobe-dominated AGN (\citealt{kellermann1988}).

In Figure 3 we show the radio intensity contours from the FIRST radio survey 
overlaid on the optical image; the radio counterpart of 2XMMJ022256.9$-$024258 is a double radio source 
having the typical Fanaroff-Riley Type II morphology 
(i.e. sharp edge lobes and bright hot spots); the absolute optical magnitude 
($M_R \simeq -24$, assuming a K-correction of $\sim 2$ mag, typical of a late type galaxy, 
see \citealt{fukugita1995})  and the total radio power 
($\sim 2.2 \times 10^{34}$ \es Hz$^{-1}$) 
are consistent with the FRII classification 
according to the dividing line in the $M_R - L_{radio}$ plane
between FRI and FRII radio galaxies (see \citealt{ghisellini2001}).

The projected separation of the two bright hot spots 
is $\sim 41$ arcsec on the sky, corresponding to a physical projected  size of
$\sim 0.33$ Mpc at the redshift of the source.
Given the discussed radio properties, the object is clearly a radio-loud 
and lobe-dominated AGN.  In this respect, 
we note that the measured size and total radio luminosity are fully consistent with that observed 
in other radio quasars (see Figure 4 in \citealt{kuzmicz2012}).

Considering the broad band properties discussed above, 2XMMJ022256.9$-$024258 can be considered 
another example of the rare class of Radio Loud Type 2 QSO, with overall properties very similar to those shown by 
e.g. AXJ0843+2942 (\citealt{dellaceca2003}), 6C0905+39 (\citealt{erlund2008}) or 4C+39.29 (\citealt{gandhi2006}). 
Indeed according to the data reported in \citet[see their Figure 3]{erlund2008} this object is one of 
the most powerful sources currently know; its intrinsic X-ray luminosity ($\sim 2\times 10^{45}$ erg s$^{-1}$), if compared with its 
178 MHz luminosity ($\sim 1.3\times 10^{35}$ erg s$^{-1}$ Hz$^{-1}$),  is more than a factor 10 above the locus sampled 
by the Narrow Line Radio Galaxies at $z<1.0$ in the 3CRR catalog and similar to that observed in Broad Line Radio Galaxies and QSOs (see Figure 3 in \citealt{hardcastle2009}). 

\item{\underline {2XMMJ100038.9+050955; \fxo=138}} 

The optical magnitude of the single object in the error circle (see Figure 2) is R=23.6. The gathered optical spectrum, unfortunately characterised by a very low SNR, seems to be rather flat and featureless, thus we have no redshift information for 2XMMJ100038.9+050955. The source is detected in all the WISE bands while there is no radio detection at the source position. 
Assuming that 2XMMJ100038.9+050955 is an absorbed QSO and using the relation between \fxo and the intrinsic 2-10 keV luminosity (\citealt{brusa2010}, see Section 5) we can estimate a redshift z$\sim$1.0 (in the range 0.6$-$1.6 taking into account a scatter of 0.5 dex in the relation, see Figure 4, left panel). At z$\simeq$1.0 the intrinsic 2-10 keV luminosity would be $\sim 9\times 10^{44}$ \es while the intrinsic $N_H \sim 5\times 10^{22}$ cm$^{-2}$. 

\item{\underline {2XMMJ121026.5+392908; \fxo=50}} 

This source was the only one not observed during our GTC run; it is a well known BL Lac object at z=0.617 
(\citealt{caccianiga2002}; \citealt{plotkin2010}). The observed optical magnitude (R=19.16), the radio 
($19\pm 0.7$ mJy at 1.4 GHz from the NVSS) and the X-ray  ($\sim 3.6\times 10^{-12}$ \ecs in the 2-10 keV band) fluxes allow us to compute multi-wavelength spectral indices
\footnote{
$\alpha_{RX}= -{log(f_{5 GHz}/f_{2 keV}) \over 7.68}$; 
$\alpha_{OX}= -{log(f_{2 keV}/f_{2500 \AA}) \over 2.605}$; 
$\alpha_{RO}= -{log(f_{5 GHz}/f_{2500 \AA}) \over 5.38}$
where $f_{5 GHz}$, $f_{2 keV}$ and $f_{2500 \AA}$ are the K-corrected fluxes at 5 GHz, 2 keV and 
$2500 \AA$, respectively.
} 
($\alpha_{RX}$ = 0.54; $\alpha_{OX}$=0.61; $\alpha_{RO}$=0.50) which are fully consistent with the BL Lac  classification (\citealt{stocke1991}; \citealt{caccianiga1999}). 
The measured $\alpha_{RX}$ ($<0.8$) suggests a high-frequency-peaked BL Lac object (HBL) in which the low-energy component of their usual, double-peak, SED peaks between the UV band and X-rays (\citealt{padovani1995}).
The observed X-ray spectrum is well described by a featureless power-law model ($\Gamma = 2.23^{+0.01}_{-0.01}$) with intrinsic $N_H = 6\times 10^{20}$ \cm and intrinsic 2-10 keV luminosity of $\sim 6.4 \times 10^{45}$ \es.
In the WISE survey 2XMMJ121026.5+392908 is detected only at 3.4 and 4.6 $\mu$m (WISE band W1 and W2, respectively).

\item{\underline {2XMMJ121134.2+390054; \fxo=50}}

The optical object clearly visible at the centre of the error circle (see Figure 2) has a magnitude R=20.77. 2XMMJ121134.2+390054 is a well known object since the epoch of the Einstein Extended Medium Sensitivity Survey (\citealt{stocke1991}) and it is classified as a BL Lac object (MS1209+3917; \citealt{rector2000}). The observed optical magnitude,  radio ($10.6\pm 0.6$ mJy at 1.4 GHz from  NVSS) and X-ray  ($\sim 8.2\times 10^{-13}$ \ecs) fluxes imply multi-wavelength spectral indices
($\alpha_{RX}$ = 0.58; $\alpha_{OX}$=0.62; $\alpha_{RO}$=0.56) fully consistent with the BL Lac classification (\citealt{stocke1991}; \citealt{caccianiga1999}). As for 2XMMJ121026.5+392908 discussed above, also 2XMMJ121134.2+390054 can be classified as a HBL object.

From the analysis of the optical spectrum shown in Figure 2 we reach the conclusion that the only clear feature is that at $\sim$ 7048 \AA, most likely associated with [OII]$\lambda$3727 \AA \ at z=0.89; we estimate an observed equivalent width (EW) of $\sim 12-15$ \AA, that rescaled to z=0 imply an EW $\sim 6-8$ \AA, very close to the limit used to classify a source as a BL Lac object (EW$<$5 \AA, \citealt{stocke1991}). 

While our spectroscospic classification as a BL Lac object agrees with previous results, the redshift proposed here (z=0.89) is significantly different from that proposed by \cite{rector2000} (z=0.602). 
These latter authors did not find any  evidence of emission line(s) in their spectrum, and based their tentative redshift determination on low SNR absorption features. We are confident that the line observed at $\sim$ 7048 \AA \ is real and this line does not have any reliable identification if the object is at z=0.602. Moreover at the proposed redshift (z=0.89) we were able to reproduce quite well also the shape of the underlying optical continuum (see Figure 2). Finally we note that this source is strongly variable: 
at least a factor of 5.5 in the X-ray domain (by comparing the measured 0.1-2.4 keV flux with the flux in the ROSAT All Sky Survey)
and a factor of 3 in the optical domain (by comparing our magnitude with the V$_{mag}$=20 reported in \citealt{rector2000}), 
so this variability could explain the clear different shape from the optical spectra reported in \cite{rector2000} and that reported here (see Figure 2).
 
At z=0.89 the observed X-ray spectrum is well described by a featureless power-law model 
($\Gamma = 2.21^{+0.07}_{-0.06}$) with intrinsic $N_H = 1.9\times 10^{21}$ \cm; the intrinsic 2-10 keV luminosity is $\sim 3.7 \times 10^{45}$ \es. The intrinsic absorption is significantly in excess to the Galactic value and this is at odds with what is usually observed in BL Lacs and, in particular, in HBL objects (see e.g. \citealt{massaro2011a} and references therein). On the other end the X-ray spectrum could be intrinsically curved due to the bump expected in HBL, peaking between the UV and X-ray bands. We therefore 
fitted the X-ray data with a broken power-law model filtered by the Galactic absorption; we find a good fit ($\chi_{\nu}^{2}$ = 1.1 for 194 d.o.f.) with best-fitting parameters $\Gamma_1 = 1.57^{+0.1}_{-0.2}$, $\Gamma_2 = 2.30^{+0.10}_{-0.13}$ and  $E_c = 1.23^{+0.2}_{-0.28}$, which are consistent with the results reported in \cite{massaro2011a} on a sample of HBL sources. The intrinsic 2-10 keV luminosity obtained with the broken power-law model is equal to that previously reported in Table 1. 
In analogy with the other confirmed BL Lac object in the sample (2XMMJ121026.5+392908), in the WISE survey this source is detected only at 3.4 and 4.6 $\mu$m . 

\item{\underline {2XMMJ123204.9+215254; \fxo=1118}} 

Two optical sources (labeled with A and B in the finding chart reported in Figure 2) are clearly present inside the 
X-ray error circle;  they have an R magnitude of R=$23.94\pm 0.41$ (source A) and R=$24.73\pm 0.64$ (source B). An optical spectrum of source A, obtained summing up two independent exposures, is shown in Figure 2; the spectrum has a clear red continuum and a significant line at $\sim 6569$ \AA. Assuming this line to be  [OII]$\lambda$3727 \AA \  at z=0.763 another observed feature is in very good agreement with being [OIII]$\lambda$5007 \AA \ ; these two features are clearly seen in both the single exposures and are not associated with strong sky lines. We have also tried to take an optical spectrum of the source B, but the source is very faint and no useful information could be extracted from the very noisy spectrum.

2XMMJ123204.9+215254 is detected in all the WISE bands discussed here. The WISE position is coincident with a source  detected in the K band (K=$18.07\pm 0.05$, see \citealt{delmoro2009}) and both are significantly closer to the brightest optical source (source A), strongly suggesting that the most probable optical counterpart of the X-ray source 2XMMJ123204.9+215254 is the source A. In the following of this paper we therefore assume that the 2XMMJ123204.9+215254 is spectroscopically identified with a Type 2 QSO at z=0.763. 

2XMMJ123204.9+215254 was previously discussed in \cite{delmoro2009} who report an infrared spectrum taken at the Subaru telescope (with the MOIRCS instrument) clearly revealing the presence of a line at the observed frame of 
$1.8837 \mu$m. Lacking optical spectroscopy and a deep optical image, these authors discussed several possible identifications for the infrared line and suggested the H$\alpha$ line at z=1.87 as its most probable origin. As discussed above our optical spectroscopy at the GTC suggests a lower redshift of z=0.763.
In this case the NIR feature reported in  \citet[observed at 1.8837 $\mu$m]{delmoro2009} could be associated with the line complex HeI$\lambda$10830+Pa$\gamma$10941, a strong and quite common  feature in AGN  (see \citealt{glikman2006}), expected to be at $\sim$ 1.92 $\mu$m (at z=0.763), slightly higher than the observed feature.
However the observed NIR line falls in a wavelength range dominated by the effect of atmospheric absorption, so 
it is difficult to determine the intrinsic line centroid; all in all, we consider the association of the infrared line with the HeI$\lambda$10830+Pa$\gamma$10941 complex as highly plausible.
On the contrary none of the faint features that we see  in the optical spectrum could be associated with any relevant emission lines from an AGN in the case of z=1.87.

At the proposed lower redshift (z=0.763) the X-ray spectrum is described by an absorbed power-law model having
a very flat photon index, $\Gamma=1.31\pm 0.23$, and an intrinsic $N_H = 3.6^{+1.03}_{-0.92}\times 10^{22}$ \cm. The intrinsic 2-10 keV luminosity is $\sim 1.9 \times 10^{45}$ \es. Since the best-fitting photon index is rather flat we test the stability of the 
measured $N_H$ and luminosity assuming a typical AGN photon index ($\Gamma=1.9$); 
we derive an intrinsic $N_H = 5.9^{+0.73}_{-0.65}\times 10^{22}$ \cm and an intrinsic 2-10 keV luminosity of $\sim 2.3 \times 10^{45}$ \es in 
good agreement with the previous values.
We note that 2XMMJ123204.9+215254 is not only the object with the highest \fxo in this sample, but (to our knowledge) it is the non-transient spectroscopically identified source with the highest \fxo discovered so far.

\item{\underline {2XMMJ135055.7+642857; \fxo=458}}

A very faint object (R=25.0) is visible at the centre of the error circle. 
Unfortunately the optical spectrum is very noisy and no features are clearly detected; we have no redshift information for this object. The X-ray spectrum is described by a power-law model with photon index  $\Gamma \sim 2$ and absorbing column density greater than $\sim 2\times 10^{21}$ \cm (obtained assuming z=0). 
An emission line is possibly detected in the EPIC-MOS spectrum (see the residuals in Figure 2, where the model does not include the line). However no compelling evidence of the presence of this line in the EPIC-pn is found, although different spectral binnings were tried.

If this line is real, and associated with the Fe K$\alpha$ emission line (the most prominent feature in the X-ray spectrum of an AGN), the implied redshift would be z$\sim$ 0.65, the absorbing column density $\sim 4\times 10^{21}$ \cm and the intrinsic 2-10 keV luminosity $\sim 2.5 \times 10^{44}$ \es; the line has an observed equivalent width of EW $\sim$ 400 eV.
Interestingly the source is also a strong (and compact) radio source, detected at 
15 GHz ($236\pm 1$ mJy, \citealt{richards2011}), 
8.4 GHz ($\sim 376$ mJy, \citealt{healey2007}) and at 
1.4 GHz ($183.5 \pm 5.5$ mJy, NVSS survey). 
The broad band spectral indices 
($\alpha_{RX}$ = 0.88; $\alpha_{OX}$=0.31; $\alpha_{RO}$=1.15), computed with the fluxes reported above, and the possible presence of a Fe K$\alpha$ line in the X-ray spectrum are not consistent with a BL Lac classification.
We remind that the X-ray source is point-like and detected almost in the center of the EPIC fields, so a high z cluster of galaxies is very unlikely. 

The radio spectrum is very flat (indeed almost inverted, with a maximum at $\sim$ 8 GHz) suggesting  that this source could be classified as 
a Giga-Hertz Peaked Spectrum radio source, a class of sources supposed to be young radio galaxies which are often characterised by 
high intrinsic absorption (see e.g. \cite{guainazzi2006}).

Assuming that this source is an absorbed QSO 
and using the relation between \fxo and the intrinsic 2-10 keV luminosity (see Section 5)  we can estimate a redshift z$\sim$1.7 (z in the range 1.1$-$2.7, inconsistent with the redshift estimated by the putative iron line), with an intrinsic 2-10 keV luminosity $\sim 3\times 10^{45}$ \es and an intrinsic $N_H \sim 1.2\times 10^{22}$ cm$^{-2}$. 
This source is detected in all the WISE bands. 

\item{\underline {2XMMJ143623.8+631726; \fxo=55}} 

A faint object with optical magnitude R=22.16 is present in the centre of the error circle (see Figure 2). 
The optical spectrum of this object is shown in Figure 2; the two marked lines can be associated with 
[OII]$\lambda3728$ and H$\gamma$+HeI (observed FWHM $<$ 1100 \kms) at z=0.893.
The observed X-ray spectrum is described by an absorbed power-law model with intrinsic $N_H = 1.46\times 10^{22}$ \cm;
the intrinsic 2-10 keV luminosity is $\sim 8.4 \times 10^{44}$ \es. The source is detected in all the WISE bands; there are no radio detections at the source position. 
We classify this source as a Type 2 QSO.

\end{itemize}

\section{Discussion}

Having defined a small but representative sample of bright  EXO50 objects, it is now instructive to compare their broad band properties with those of other samples of AGN, both absorbed and unabsorbed, from the literature. 

\subsection {The comparison sample}

We have assembled an heterogeneous sample of spectroscopically confirmed X-ray selected Type 1 and Type 2 AGN from  a few surveys carried out during the last few years using XMM-Newton data: the COSMOS survey (\citealt{brusa2010}), the XMS survey (\citealt{barcons2007}, \citealt{carrera2007}) and the XBS survey (\citealt{dellaceca2004}; \citealt{caccianiga2008}). This first large comparison sample, composed by 882 Type 1 AGN and 487 Type 2 AGN,  will be used below (see Figure 4, left panel) to discuss the position of absorbed and unabsorbed AGN in the \fxo vs. $L_X$ plane.

As a second step, and in order to compare the optical-infrared photometric properties of the EXO50 objects with objects having similar luminosities and redshifts, we have considered in the comparison sample only the high luminosity (intrinsic $L_{(2-10\ keV)} > 10^{44}$ \es) sources (i.e. the Type 1 QSO and Type 2 QSO) in the redshift range between 0.6 and 1.1; we have cross-correlated this latter sample of sources with the WISE All-Sky source catalogue, using a searching radius equal to 2 arcsec (consistently with the results obtained between the cross-correlation of the EXO50 sources and the WISE catalog, see Section 3.3).

The WISE-QSO sample to be used in the comparisons depends also on the WISE band(s) considered; to obtain meaningful colour distributions we use a similar approach followed by  \cite{yan2013} and base our comparisons (see below)  on the following two WISE-X-ray selected QSO samples: the QSO W1/2 sample ($S/N_{W1} \geq 7$ and $S/N_{W2} \geq 5$; 94 Type 1 QSO and 11 Type 2 QSO), the QSO W1/2/3 sample ($S/N_{W1} \geq 7$, $S/N_{W2} \geq 5$ and $S/N_{W3} \geq 3$; 72 Type 1 QSO and 7 Type 2 QSO); the QSO W1/2/3 sample is obviously a subset of 
the QSO W1/2 sample

\subsection {The X-ray to optical flux ratio vs Lx and the optical-IR colour}

\begin{figure*}
\label{fig}  
\centering
\subfigure{ 
  \includegraphics[height=8.0cm, width=8.0cm,angle=0]{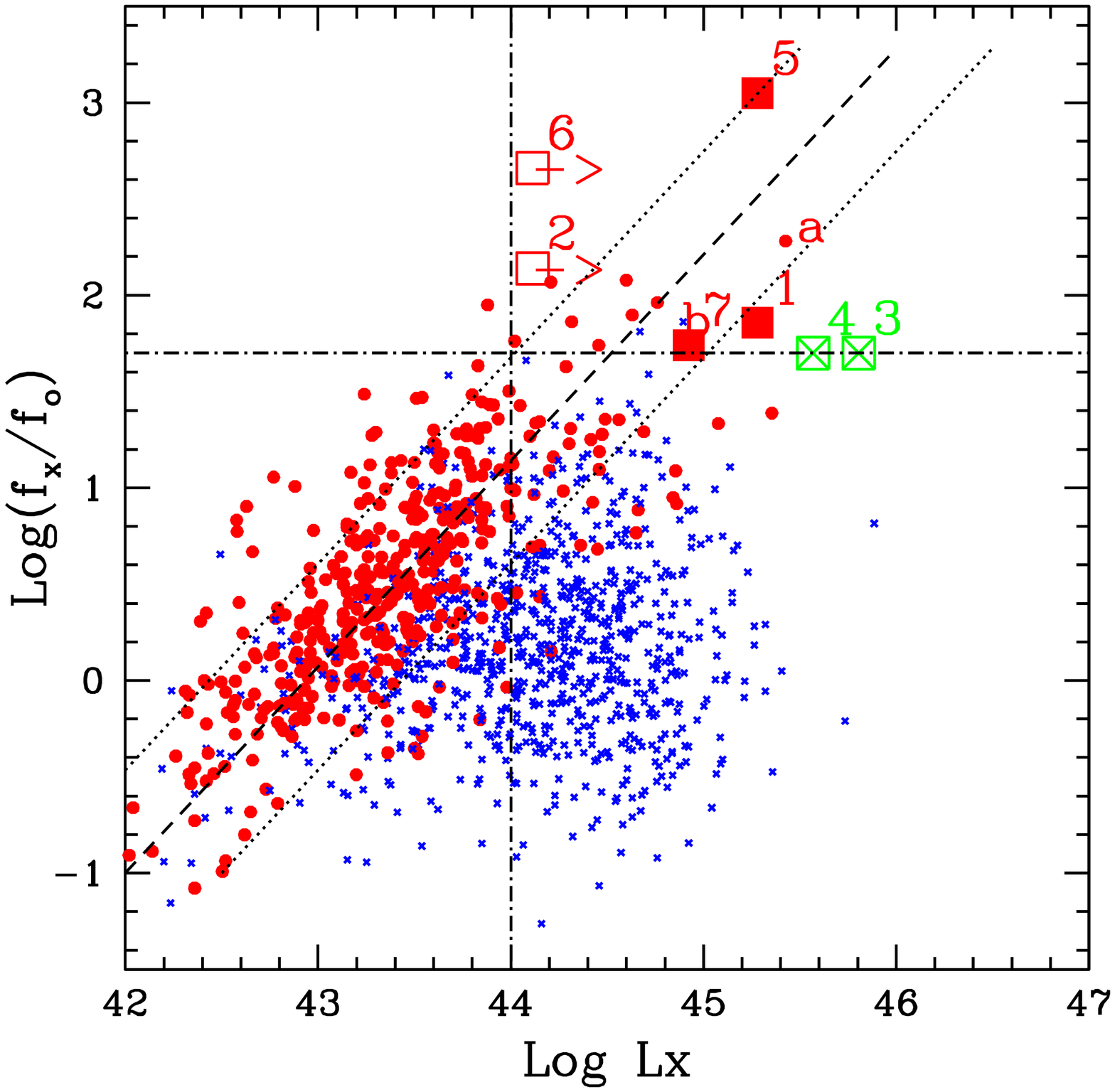}
  \includegraphics[height=8.0cm, width=8.0cm,angle=0]{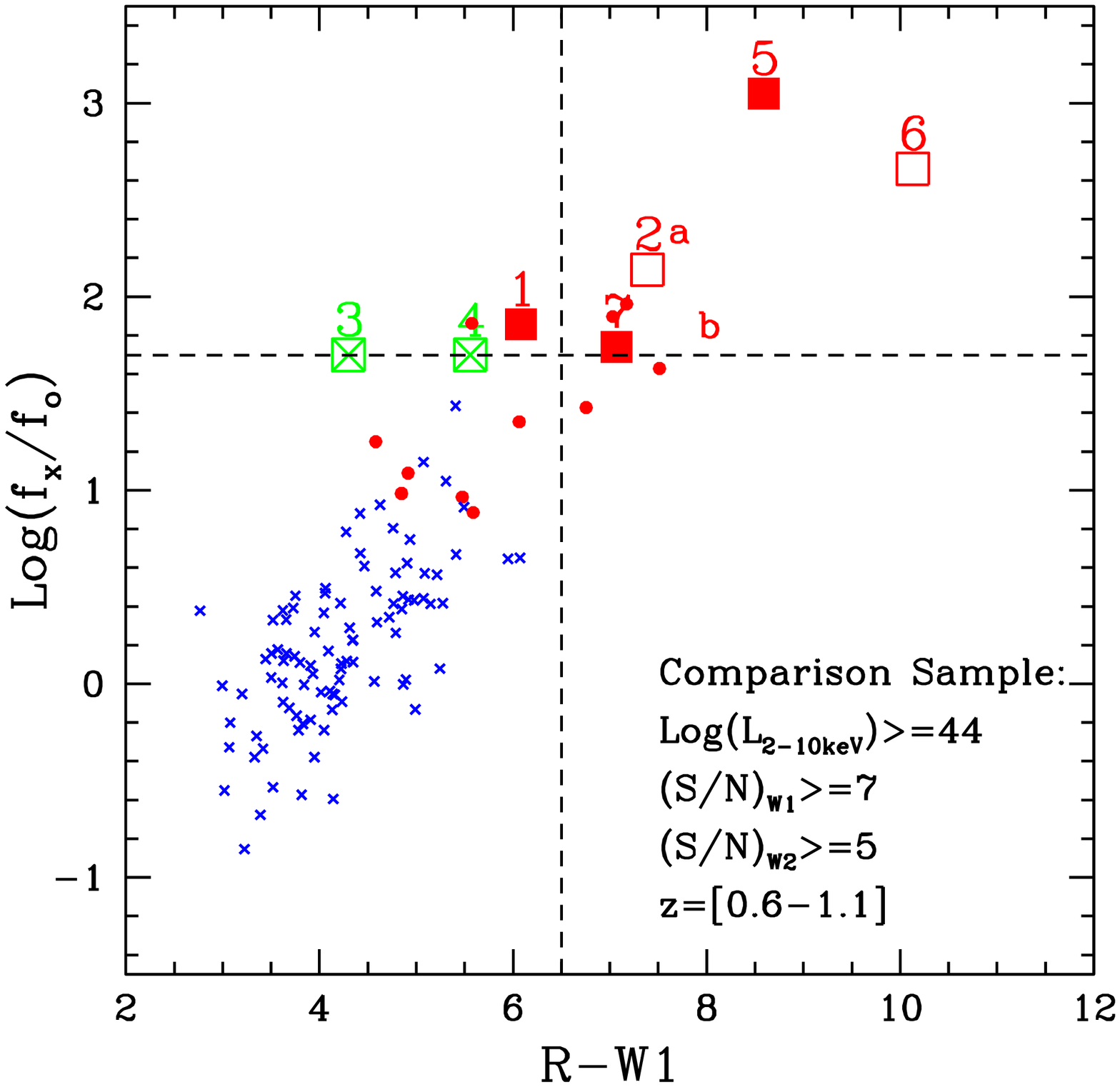}}      
\caption{{\it Left panel}: relationship between the measured  \fxo (in logarithmic units)  and the intrinsic 2-10 keV luminosity for our sample of EXO50 objects and for the comparison sample of AGN assembled as described in Section 5.1. 
Large squares: our sample of EXO50 objects (filled-red for Type 2 QSO, open-red for 
unidentified objects  and 
open-crossed-green for BL Lac objects)
the numbers close to the large squares  mark the EXO50 objects as reported in Table 1 and Table 3.
The labels a) and b) mark the position of the objects 
XBSJ021642.3-043553 (z$\sim2$, source a) and XID2028 (z$\sim1.6$, source b) discussed in section 5.2.
Filled red circles: Type 2 AGN from the comparison sample; 
Small blue crosses: Type 1 AGN from the comparison sample;
The dashed diagonal corresponds to the relation between \fxo and the intrinsic 2-10 keV luminosity for 
obscured AGN, while the two dotted lines corresponds to a scatter of about 0.5 dex around this 
relation. 
{\it Right panel}: 
the \fxo ratio as a function of the optical-infrared color R-W1 (a proxy of the usual R-K colour) for the EXO50 objects and for the sample of QSO W1/W2 as described in the text; the vertical dashed line  (R-W1=6.5) corresponds roughly to R-K=5.
The horizontal dashed line correspond to \fxo = 50. Symbols and colours are as in the left panel.} 
\end{figure*}

In Figure 4, left panel, we report the EXO50 objects discussed in this paper and the comparison sample of Type 1 and Type 2 AGN from the literature.
In the comparison sample 12 objects have \fxo $>$ 50 and all but two
\footnote{These two sources, classified as broad line AGN, are 
CL0016\_3 (z=1.09) from the XMS survey and XMMC\_150.13303+2.30324 (z=1.6) from the COSMOS survey.} 
have been classified as Type 2 QSO. Unfortunately the large majority of these 12 sources (some of them will be highlighted in some figures reported below) 
are faint both in X-ray and in the optical domain, so only a couple of these interesting objects have been discussed in some detail in the literature. 
In particular we have the Type 2 QSO XBS~J021642.3-043553 (z=1.98; \fxo $\sim$ 200; $L_{(2-10 keV)} \sim 3\times 10^{45}$ \es, see \citealt{severgnini2006}) and the Type 2 QSO XMMC\_150.54703+1.61869 (XID2028; z=1.59; \fxo $\sim$ 60; $L_{(2-10 keV)} \sim 8\times 10^{44}$ \es, see \citealt{brusa2010}, \citealt{brusa2015}); both objects are considered as prototypes of the obscured QSO population at high redshift, where the coexistence between massive galaxies ($\sim$ few times $10^{11}\ M_{\odot}$ as estimated in the above papers) and powerful QSOs has been proved. Furthermore, in the case of XMMC\_150.54703+1.61869, a recent paper (\citealt{perna2014}) reports the discovery of a massive outflow ($> 630\ M_{\odot}\ yr^{-1}$), extending out to 
10 Kpc from the central black hole and having a total energetic in full agreement with the prediction by 
AGN feedback models.

The dashed diagonal line in Figure 4 (left panel) corresponds to the relation between \fxo and the intrinsic 2-10 keV luminosity for obscured AGN initially proposed in \cite{fiore1999} and revised by \cite{brusa2010}, while the two dotted lines corresponds to a scatter of about 0.5 dex around this 
relation. As discussed in the previous section, 3 EXO50 objects are spectroscopically identified as Type 2 QSOs (the sources number 1, 5 and 7 in the figure); taking into consideration the scatter that we see for the other Type 2 AGN these 3 EXO50 objects seems to follow the same relationship between \fxo and the intrinsic 2-10 keV luminosity. 

It is worth noting the position of the two EXO50 objects spectroscopically identified as BL Lac objects (the sources number 3 and 4); both objects seem to have a larger X-ray luminosity for a given \fxo if compared with the absorbed AGN population. 
This is indeed expected, since BL Lacs have SED strongly dominated by the beamed emission from the jets. 
Their high \fxo therefore is not due to obscuration (as for the Type 2 QSO where the optical luminosity is very likely dominated by the host galaxy) but to the intrinsic shape of the SED (see also below).

In Figure 4 (right panel) we plot  the \fxo ratio as a function of the optical-infrared color (R-W1) for our EXO50 objects and for the QSO W1/W2 sample defined in Section 5.1.The colour  R-W1 can be considered a good proxy for the usual R-K colour and the vertical dashed line  (R-W1=6.5) corresponds
\footnote {This correspondence has been derived from the analysis (not reported here) of the sources in the COSMOS survey where we have both R, K and W1 magnitudes.}  roughly to R-K=5, the value used to define an object as a Extremely Red Object (ERO, see e.g. \citealt{elston1988}, \citealt{daddi2000}). X-ray emitting EROs studied so far, strongly suggest that the bulk of this population is composed by obscured AGN (see e.g.  \citealt{mignoli2004}, \citealt{severgnini2005}, \citealt{brusa2005}) and the results obtained here support this view
\footnote {We note that both objects from literature discussed in the previous section are EROs (R-K$\geq$5 for  XBS~J021642.3-043553 and R-K=6.46 for XMMC\_150.54703+1.61869).}.

A clear trend between the measured \fxo and the optical-infrared colours is present. This trend, visible both for the Type 1 QSO and for the Type 2 QSO samples (see also  \citealt{brusa2005}), could be 
partially explained as due to the effect of absorption (see also Section 5.6).
As previously said, the nuclear optical fluxes are more depressed (indeed completely blocked in the case of Type 2 QSO where we should see only the host galaxy) by circum-nuclear matter if compared with the infrared and X-ray fluxes.
As already noted in \cite{caccianiga2011} the effect of absorption is important also in Type 1 AGN and this explain 
why we see the trend also in Type 1 QSOs. A further support to this hypothesis comes from the observation that, for the Type 1 QSOs in Figure 4 (right panel) belonging to the XBS survey (for which we have X-ray and optical spectral information), there is a clear evidence that the sources with the higher \fxo and redder colours are also those with the higher intrinsic $N_H$. 

Two out of three of the EXO50 sources classified as Type 2 QSO  are on the "EROs" side of the diagram (R-W1 $>$ 6.5); the only object just outside the EROS locus is 2XMMJ2256.9$-$024258 (the EXO50 object at z$\sim$1 associated with the double and bright  radio source 4C-02.11), although its position in the plot is not significantly different from the other sources considering the scatter on the \fxo vs R-W1 relation.
The position of the two unidentified EXO50 in the \fxo vs. R-W1 plot (Figure 4, right panel, open squares), in the R-W2 vs. W1-W2 plot (Figure 5, open squares) and in the \fxo vs. R-W2 plot (Figure 7, open squares)  strongly suggest an obscured QSO nature; if we assume that the two unidentified sources follow the \fxo $-L_x$ relation, their intrinsic $L_x$ would be  $2.5\times 10^{44} - 2.5\times 10^{45}$ \es (which would imply a redshift in the range $0.6-1.6$) for 2XMMJ100038.9+050955 and $9\times 10^{44} - 9\times 10^{45}$ \es (which would imply a redshift in the range $1.1-2.7$) for 2XMMJ135055.7+642857.

\subsection {The optical-IR flux ratio vs the IR colours}

The optical-infrared colours (R-W2) for our EXO50 objects and for the comparison QSO W1/2 sample as a function of the infrared colour W1-W2 are shown in Figure 5.
We have also marked the place (W1-W2$>$0.8 and R-W2$>6$) where, according to \cite{yan2013} , it is  possible to select Type 2 QSO candidates.
The two EXO50 objects spectroscopically identified as BL Lac objects seem to have optical-IR properties different from the bulk of the AGN population (see also \citealt{massaro2011b} for similar results based on WISE MIR colours).
We stress that our EXO50-Type 2 QSO objects are amongst the most extreme sources in these diagrams (see Figure 15 in \citealt{yan2013} for a comparison with the SDSS QSO sample).  

As can be seen, all the QSOs, regardless of their spectroscopic or \fxo  properties, span a very similar range of the infrared (W1-W2) colour, although there are a couple of our EXO50 objects with extreme W1-W2 colours (source 2 and source 5, namely 2XMMJ100038.9+050955 and 2XMMJ123204.9+215254).
On the contrary the  optical-infrared colours (R-W2 or R-W1, see also Figure 4, right panel) have to be in some way 
correlated with the X-ray to optical flux ratio properties, and with the optical spectroscopic classification (see also the results 
reported in Figure 7).

The observed quite similar range of the W1-W2 colours in Type 1 and Type 2 QSOs (in agreement with similar results previously reported in e.g. \citealt{mateos2013})
suggests that the absorption is probably not the only (and the more important) parameter in shaping the AGN infrared continuum (between $\sim$1.7 and 3$\mu$m rest frame 
assuming an average redshift of 0.8).
On the contrary, the observed separation of the different classes of AGN in the R-W2 colour clearly indicates  that the absorption could be an important parameter in shaping the optical-infared observed properties. We will discuss these topics in Section 5.6. 

\begin{figure}
\begin{center}
\resizebox{0.46\textwidth}{!}{
\rotatebox{0}{
\includegraphics{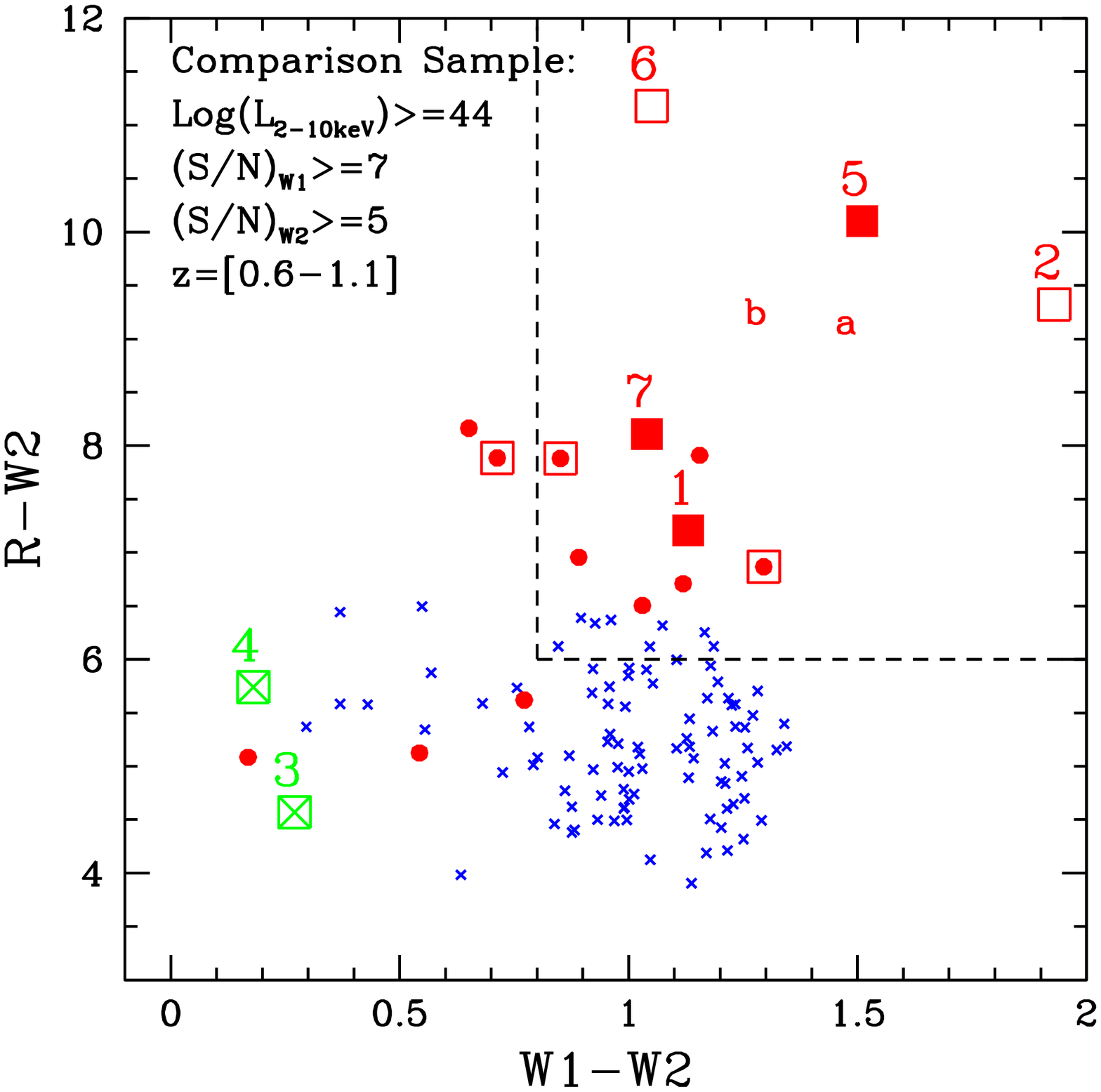}
}}
\caption{The optical-infrared  R-W2 colour vs. the infrared colour W1-W2 for the EXO50 objects and for the comparison QSO W1/2 sample (see Section 5). The zone in the plot enclosed by the dashed line (R-W2$>$6 and W1-W2$>$0.8) corresponds to the place where it is  possible to select Type2 QSO candidates (\citealt{yan2013}).
Symbols and colours are as in Figure 4; we have also marked the sources with \fxo $>$ 50 in the comparison QSO samples using empty squares. The numbers mark our EXO50 objects as reported in Table 1 and Table 3.
The labels a) and b) mark the position of the objects 
XBSJ021642.3-043553 (z$\sim2$, source a) and XID2028 (z$\sim1.6$, source b) discussed in section 5.2.
}
\label{fig}
\end{center}
\end{figure}

\subsection {The IR colours diagnostic plot}

\begin{figure}
\begin{center}
\resizebox{0.46\textwidth}{!}{
\rotatebox{0}{
\includegraphics{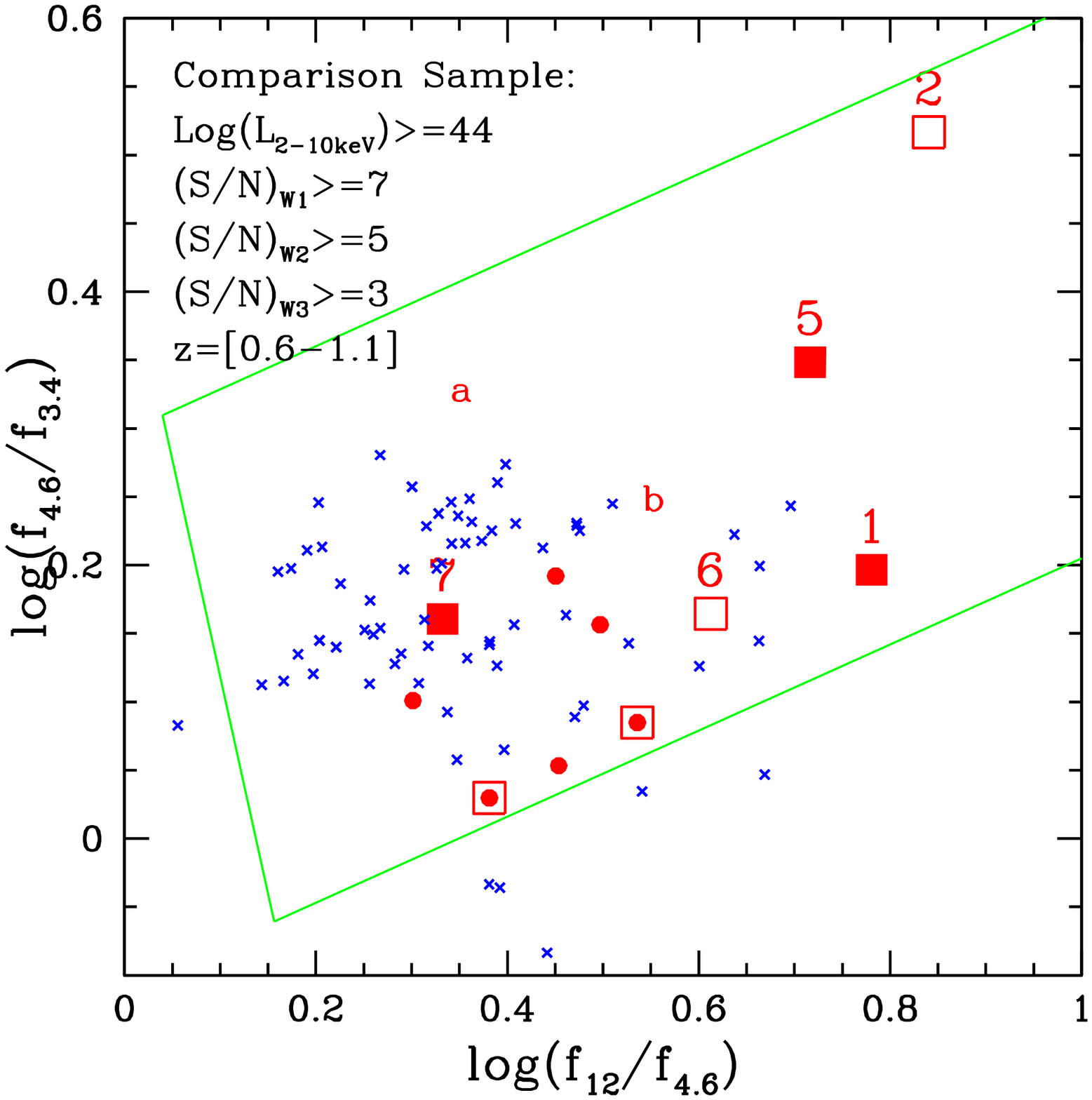}
}}
\caption{WISE MIR colours for the EXO50 objects and for the comparison QSO W1/2/3 sample 
(a subset of the QSO W1/W2 sample reported in Figure 5, see Section 5.1 for details). 
Note that the two BL Lacs are not reported here since they are not detected 
at 12$\mu$m. The solid wedge represents the AGN locus according to 
\citealt{mateos2012} (see  Section 5 details). Symbols and colours are as in Figure 4; we have also marked the sources with \fxo $>$ 50 in the comparison QSO samples using empty squares. The numbers mark the EXO50 objects as reported in Table 1 and Table 3.
The labels a) and b) mark the position of the objects 
XBSJ021642.3-043553 (z$\sim2$, source a) and XID2028 (z$\sim1.6$, source b) discussed in section 5.2.
}
\label{fig}
\end{center}
\end{figure}

In Figure 6 we show the WISE MIR colours (this time as a ratio between fluxes)  for our EXO50 objects and for the QSO W1/2/3 sample defined above; the two EXO50 BL Lac objects are not reported here, since they are not detected at 12$\mu$m. The solid wedge represents the locus populated by luminous AGN according to \cite{mateos2012}. 
As shown in \cite{mateos2012} and confirmed here, we see that the technique is highly effective at identifying both Type 1 and Type 2 QSOs. 
The position of the different kinds of QSOs in the WISE MIR colour plot does not seem to suggest  a big difference between absorbed and unabsorbed AGN in the MIR domain, 
in good agreement with recent results reported in \cite{mateos2013}. 

It is worth noting that three EXO50 objects clearly stand out from the zone populated by the bulk of AGN; two of these objects are spectroscopically identified and, interestingly,  are the sources with the highest X-ray luminosity (namely 2XMMJ123204.9+215254 and 2XMMJ022256.9$-$024258).  

\subsection {The relation between EXO50 objects and EDOGs}

The three EXO50 sources on the top-right zone of Figure 5 (the sources 2, 5 and 6) have R-W4 above 15 (see Table 1 and 3), implying a  
$f_{24 \mum}/f_{R}$ well in excess of  $\sim 3000$. 
These 3 sources can be classified as Extreme Dust Obscured Galaxies (EDOGs; $f_{24 \mum}/f_{R} \geq 2000$; \citealt{lanzuisi2009}), i.e. extreme examples of Dust Obscured Galaxies (DOG, $f_{24 \mum}/f_{R} \geq 1000$ see 
\citealt{dey2008}, \citealt{fiore2008}),
a class of galaxies firstly discovered with the Spitzer Space Telescope.
In these objects, the high infrared-to-optical ratios imply that large amounts of dust are absorbing the optical light and re-emitting it in the infrared. 
DOGs and EDOGs may play an important role in the formation of the most massive galaxies in the local Universe and, in particular,  represent an important evolutionary step in the AGN-galaxy connection (see \citealt{lanzuisi2009} and reference therein).

A sample of 44 EDOGs from the SWIRE survey having X-ray coverage (with XMM-Newton or Chandra) has been selected and studied by \cite{lanzuisi2009}; the source redshifts (both spectroscopic and photometric), are in the range between $\sim$0.7 and $\sim$2.5. About 
95\% of the detected sources
\footnote{An inadequate X-ray coverage is probably the cause of the non-detection of the remaining ones.} 
(23 objects) are consistent with being obscured by neutral gas with an intrinsic
column density (mostly derived using X-ray hardness ratios)  typically in the range between $10^{22}$ to few times $10^{23}$ cm$^{-2}$. Their intrinsic 2-10  keV luminosities (in excess to $10^{43}$ \es)
fall well within the AGN X-ray luminosity range and $\sim$ 55\% of them can be classified as Type 2 quasars, on the basis of their absorption properties and X-ray luminosity. 
It is also worth noting that, among the EXO50 sources defined here, about 40\% can be classified as EDOGs ($\sim 60\%$ excluding the BL Lac objects). 
Unfortunately, two of our EXO50 EDOGs are still unidentified, preventing us to make a proper comparison with the work of  \cite{lanzuisi2009}.
We can only say that the absorption properties of the only identified EXO50 EDOG (source number 5) are fully  consistent with the main results obtained from \cite{lanzuisi2009}: the source is an absorbed QSO with an intrinsic absorbing column density 
$\sim 3\times 10^{22}$ cm$^{-2}$. For the remaining two the Type 2 QSO hypothesis is very likely. 
 
\subsection {Are the EXO50 Type 2 QSO objects different from standard X-ray selected Type 2 QSO?}

Overall the EXO50 Type 2 QSO do not seem to be different from standard X-ray selected Type 2 QSOs in terms of nuclear absorption (in the range of 
few times $10^{22}$ cm$^{-2}$ up to $8\times 10^{23}$ cm$^{-2}$), so  other factors 
may play a possible role in explaining their extreme properties. 

\begin{figure}
\label{fig}  
\centering
\subfigure{  
  \includegraphics[height=8cm, width=8cm,angle=0]{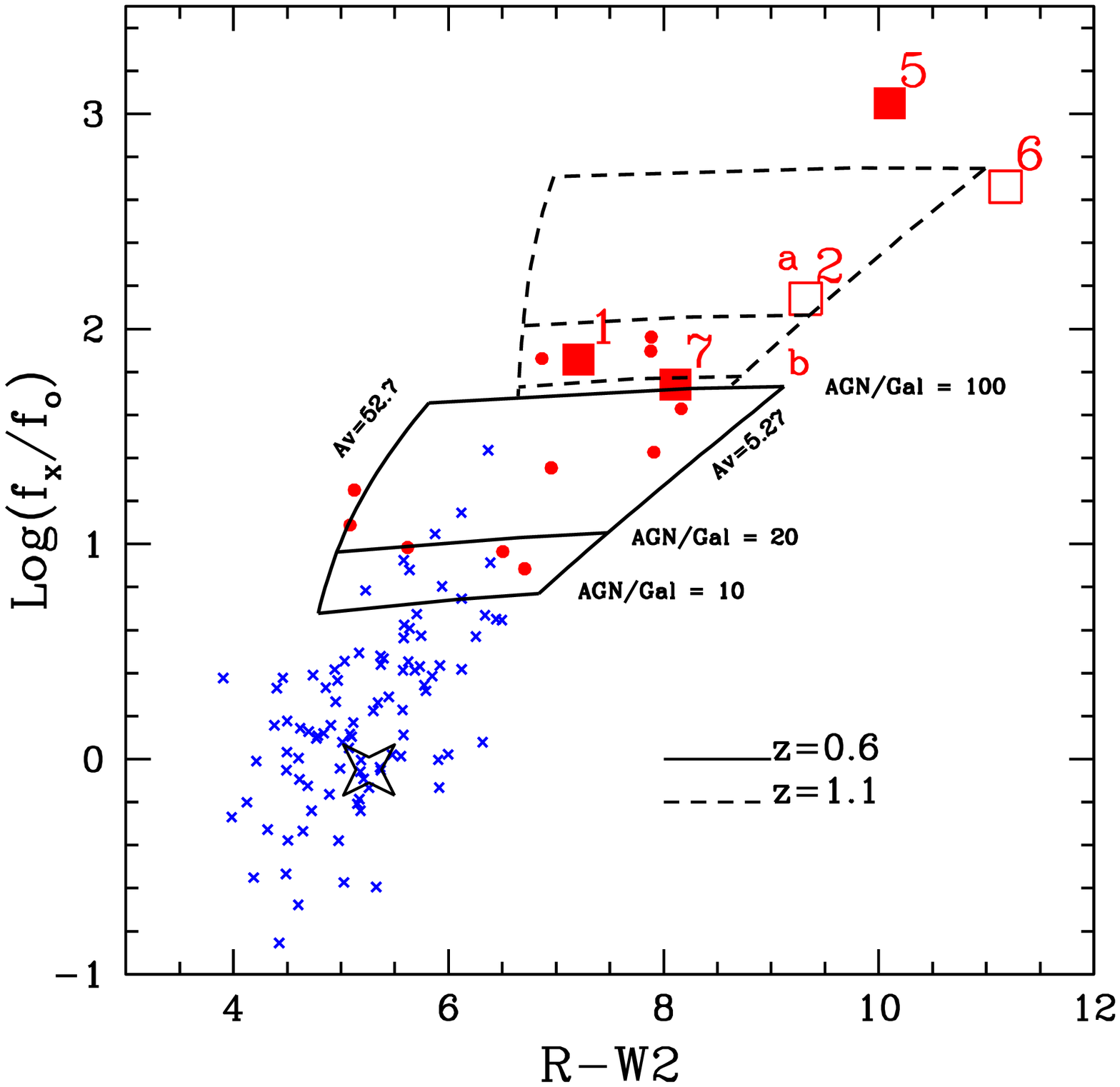}}   
\caption{The \fxo ratio as a function of the optical-infrared color for our EXO50 Type 2 QSO objects and for the comparison sample. We also report the expectations from the {\it toy model} at z=0.6 and z=1.1 for two values of optical obscuration (Av=5.27 and 52.7 corresponding to $N_H \simeq 10^{22}$ and $10^{23}$ cm$^{-2}$, respectively) and three values (10, 20, 100) of AGN/host ratio at $\sim$ 4.6 $\mu m$
; the big star marks the position where the {\it toy model} converges for all the AGN/host ratio used here when 
$N_H <$ few times  $10^{20}$ cm$^{-2}$ (i.e. at the unabsorbed Type 1 view). Symbols and colours are as in Figure 4. The numbers mark our EXO50 objects as reported in Table 1 and Table 3. 
The labels a) and b) mark the position of the objects 
XBSJ021642.3-043553 (z$\sim2$, source a) and XID2028 (z$\sim1.6$, source b) discussed in section 5.2.
See Section 5.6 for details on the modelling.
}
\end{figure}

To investigate how the\fxo  and the optical-to-infrared colours change as a function of a) the intrinsic absorption of the AGN component; b) the luminosity contrast between the intrinsic AGN and the host galaxy; and c) the redshift, we have used a very simple {\it toy model} to describe the SED of the AGN+host galaxy complex. 
 
As unobscured AGN SED we have used the Type 1 QSO UV-to-IR template described in \cite{polletta2007} adding the X-ray component assuming a \fxo (as defined in Section 2) $\simeq$  1 (typically observed in unabsorbed AGN),  
while for the host galaxy we have used a Sc galaxy template; similar trends to those discussed below are obtained if we assume a different morphology for the host galaxy template.
Full details on the AGN and host SED as well as on the procedure summarised below can be found in \cite{ballo2014}.

The UV-to-IR AGN template is absorbed at the source redshift using the extinction curve of the Galactic center (\citealt{chiar2006}) before summing it up  with the galaxy template; the X-ray absorption is tied to the optical obscuration assuming a Galactic gas to dust ratio ($N_H$/Av $= 1.9\times 10^{21}$ cm$^{-2}$, \citealt{bohlin1978}). The variables of this simple {\it toy model} are thus the redshift, the AGN dust extinction (and consequently the X-ray absorption) and the ratio between the {\bf intrinsic} AGN emission and the host galaxy emission (normalised for our convenience at $\sim$ 4.6 $\mu m$, observed frame).

In Figure 7 we show the expectations of this {\it toy model} in the \fxo vs. R-W2 plane at a redshift of 0.6 (solid line) and 1.1 (dashed line, respectively the lower and the upper envelope of the investigated redshift range) for two values of dust extinction (Av=5.27 and 52.7, corresponding to $N_H \simeq 10^{22}$ and $10^{23}$ cm$^{-2}$, respectively) and for three values of the AGN to host galaxy ratio (10, 20 and 100). In the figure we have also reported the position of the sources belonging to the QSO W1/2 sample as well as our Type 2 QSO EXO50 sources. 
The big star marks the position where the {\it toy model} converges, for all the AGN/host ratios used here, when $N_H <$ few times $10^{20}$ cm$^{-2}$. As can be seen, the expected position of the unabsorbed AGN is fully consistent with the data for the Type 1 QSO in the comparison sample. 

Figure 7 clearly suggests that several factors may concur in explaining the extreme properties of the EXO50 Type 2 QSO. For a fixed absorption or AGN/host ratio, the higher is the redshift, the higher is the \fxo and the redder are the optical-to-infrared colours (cf. \citealt{fiore2003}). However from a deeper look to the data, this {\it redshift effect} alone is not able to explain the extreme properties of all the EXO50 objects; for example the source number 5 
(namely 2XMMJ123204.9+215254 at z=0.76) can not be explained with the {\it redshift effect}  but requires very high AGN/host ratios; a very high AGN/host galaxy ratio indeed may be required also for 2XMMJ100038.9+050955 and 
2XMMJ135055.7+642857 (sources 2 and 6, respectively) if their redshifts turn out to be similar to that of 2XMMJ123204.9+215254. For comparison we note that the 8 Type 2 QSO with \fxo in the range between 5 and 15 reported in Figure 7 belong to the XBS survey and have been studied in detail by \cite{ballo2014}; these sources have an AGN to host galaxy ratio at 4.6 $\mu m$ (observed frame) between a few and $\simeq$ 20.

A wide range in the intrinsic luminosity ratio between the AGN and the host galaxy could also explain the 
observed wide range of the W1-W2 colours in Type 1 QSO and Type 2 QSO (see Section 5.3), with the AGN component increasing, on average, going from the left (W1-W2=0, corresponding to $\alpha \simeq -$2) to the right (W1-W2=2, corresponding to $\alpha \simeq$4.3) of the abscissa in Figure 5. 
Finally, as already discussed in the previous section, the sources with the highest \fxo are also the sources with extreme optical-IR properties. 
Therefore a  high AGN/host ratio (along with the large amount of dust absorption)
could be a natural explanation to account for the extreme properties for a part of the EDOG population. 
This also implies that the infrared luminosities measured in these objects are mainly due to the AGN component rather than to starburst activity.

\subsection {The BL Lacs in the EXO50 sample}

Finally, although the statistics of EXO50 objects is very limited, it is worth noting that $\sim 30$ per cent of the EXO50 sources in the bright flux regime explored by our survey is represented by BL Lac objects. 
At the 2-10 keV flux limit  of $\sim 1.5\times 10^{-13}$ \ecs the derived surface density is $\sim 3\times 10^{-2}$ deg$^{-2}$. 
This surface density is a lower limit to the real surface density of BL Lac objects, since we are considering here only the BL Lacs with \fxo $> 50$, and corresponds to a fraction between $\sim$ 8\% and $\sim$ 30\% of the density of X-ray selected BL Lacs at similar fluxes 
(the exact fraction depending on their cosmological evolution properties, e.g. \citealt{wolter1991}).
Their position in the several plots discussed here deviates significantly from the rest of the sample.
The very high \fxo selection used here allows us to pick up the most extreme BL Lacs (HBL type) in the $\alpha_{OX}$ - $\alpha_{RX}$ plane (see also \citealt{costamante2001}). 
This is clearly evident in Figure 8, where we show the $\alpha_{OX}$-$\alpha_{Rx}$ plane for the BL Lacs objects in the BZCAT catalog (\citealt{massaro2009}) with the position of the two EXO50 sources identified as BL Lacs overlaid; 
only less than 1 percent of the BL Lacs discovered so far have broad band spectral indices as extreme as the ones discovered here using the \fxo$>50$ selection.  

\begin{figure}
\label{fig}  
\centering
\subfigure{  
  \includegraphics[height=8cm, width=8cm,angle=0]{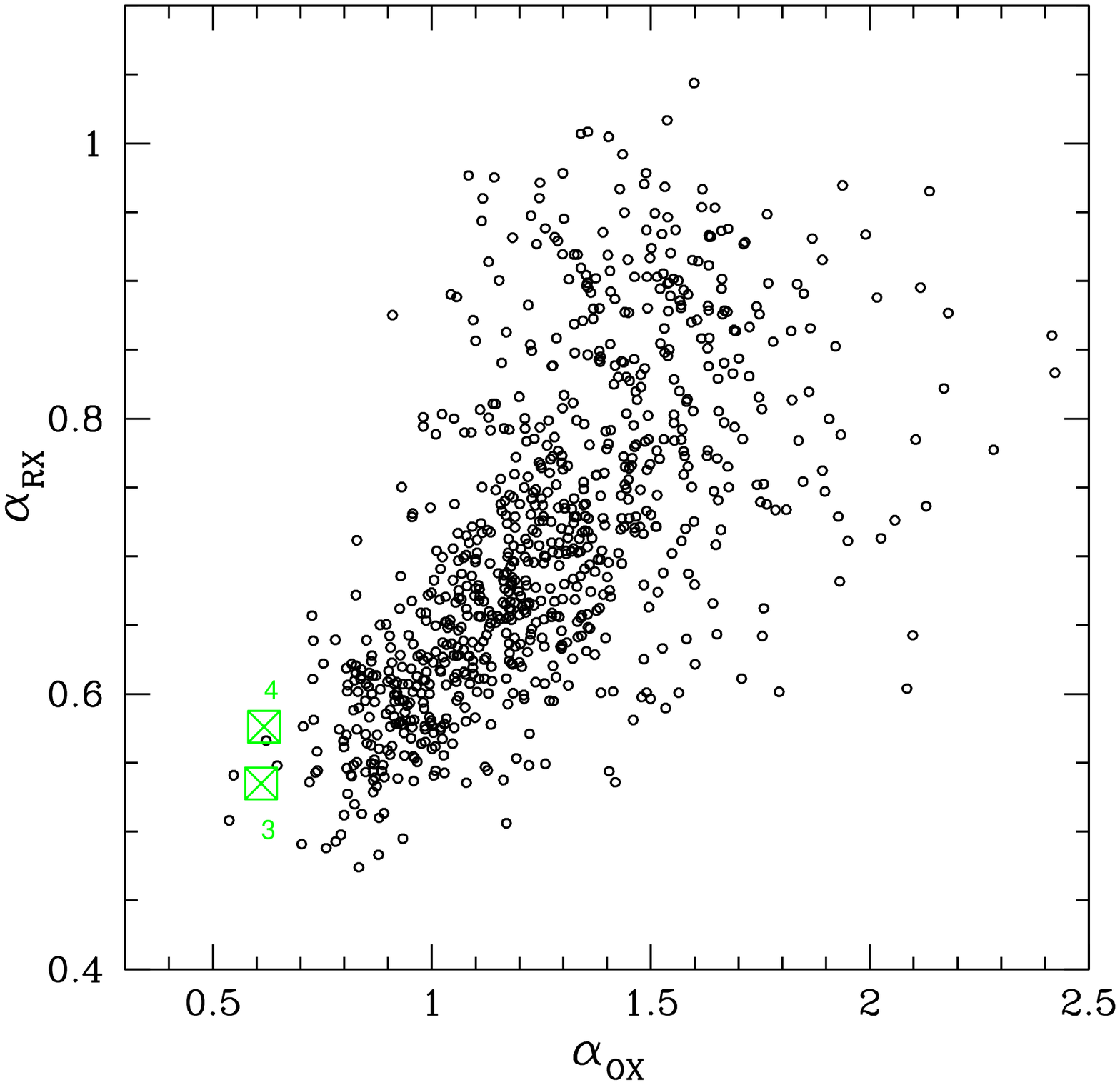}}   
\caption{The $\alpha_{OX}$-$\alpha_{Rx}$ plane for the BL Lacs objects in the BZCAT catalog (black open circle) and for the two EXO50 sources (open-crosses-green symbols) spectroscopically identified 
as BL Lacs reported in this paper 
}
\end{figure}

BL Lac objects are sources emitting non-thermal radiation across the entire electromagnetic spectrum from a relativistic jet
that is viewed closely along the line of sight, thus causing strong relativistic amplification (e.g. \citealt{urry1995}). Their SED is characterised  by two broad peaks, now almost commonly interpreted as due to synchrotron and inverse Compton radiation.  In this respect, the BL Lac objects reported here should have the synchrotron peak emission around 0.1 keV (see Figure 1 in \citealt{costamante2001}), indicating the presence of high energy electrons: these extreme BL Lacs are good candidates for TeV emission 
that, if detected at these redshifts, could provide important constraints both on the physical process at work in such sources, and on the 
intensity and spectrum of the diffuse Extragalactic Background Light (see e.g. \citealt{costamante2013}).
Presently (end of July 2014) there are about 50 blazars detected at TeV energies 
 \footnote{see http://www.asdc.asi.it/tgevcat/index.php}. Two out of the 6 detected blazars at z$>$ 0.3 are HBL (with z=0.34 and z$>$0.6) 
like the ones discussed here; the X-ray and radio fluxes of these two high-z TeV detected HBL are comparable with the fluxes of our sources and, therefore, we expect that these objects are potentially detectable at TeV energies even with the facilities currently available.
    
\section{Summary and Conclusions}

Obscured QSOs are expected to display large values of X-ray to optical flux ratio, \fxo,  where $f_x$ refers to the observed 2-10 keV flux (corrected for Galactic absorption)
and $f_o$ is the optical flux derived from the R magnitude.
In this paper we have defined a small, but  statistically complete and representative sample of 7 point-like X-ray sources characterised by an extreme X-ray to optical flux ratio, \fxo{} $>$50 (i.e more than 15 times the average values of unobscured broad line and soft X-ray selected AGN), and with $f_x>1.5\times 10^{-13}$ \ecs. The sky coverage investigated to find out these 7 objects is about 60.4 sq deg, implying a density of 
$\sim 1.2$ EXO50 sources/deg$^{2}$ at the sampled fluxes. We have spectroscopically identified (using the GTC data as well data from the literature) 5 EXO50 objects; two sources remain still unidentified, although the data at other wavelengths discussed here strongly suggest an obscured QSO nature. 

We have discussed in detail the broad band properties of each EXO50 source in our sample, and we have compared them with those of a comparison sample of X-ray selected AGN/QSO from the literature.

The main results of this investigation are the following:  

a) about 70 percent of the EXO50 sources in the bright flux regime explored by our survey are associated with obscured AGN ($N_H > 10^{22}$ cm$^{-2}$), spanning a redshift range between 0.75 and 1 and characterised by 2-10 keV intrinsic luminosities in the QSO regime (e.g. well in excess to $10^{44}$ \es). This result confirms the suggestion that the \fxo ratio can be  used as a proxy of obscuring material and it is a very efficient tool for the selection of obscured QSOs (see also \citealt{lanzuisi2013}).
If compared with other samples of Type 1 and Type 2 QSO, the EXO50- Type 2 QSO objects are amongst the most extreme sources in the several optical-infrared diagrams investigated. Overall the EXO50 Type 2 QSO do not seem to be different from standard X-ray selected Type 2 QSOs in terms of nuclear absorption; a very high AGN/host galaxy ratio seems to play a major role in explaining their extreme properties. Interestingly, 3 out of 5 EXO50 Type 2 QSO objects can be classified as EDOGs ($f_{24 \mum}/f_{R} \geq 2000$), a source population that may play an important role in the formation of the most massive galaxies in the local Universe. Similarly to what we have found for the EXO50 Type 2 QSOs, we suggest that a very high AGN/host ratios (along with the large amount of dust absorption) could be the natural explanation for the observed properties for part of the EDOG population.
Two recent papers (\citealt{brusa2015}, \citealt{perna2014}) find the presence of massive outflows in objects similar to the ones discussed here. Unfortunately the optical spectra presented in this paper does not have the quality required to study in any detail the presence of outflows; an approved LBT programme on this topic is ongoing;

b) the remaining EXO50 sources are represented by BL Lac objects. 
Interestingly the very high \fxo selection used here allows us to pick up rather extreme BL Lacs, which are good candidates for TeV emission. If detected, they could provide important constraints both on the physical process at work, and on the 
intensity and spectrum of the diffuse Extragalactic Background Light.

\section*{Acknowledgments}

This research has made use of data obtained from the XMM-{\it Newton} 
satellite and data obtained from the High Energy Astrophysics Science
Archive Research Center (HEASARC), provided by NASA's  Goddard Space Flight Center.
We acknowledge partial financial support from ASI grants (n. I/023/05/0, n. I/088/06/ and I/037/12/0) and 
from the Italian Ministry of Education, Universities and Research 
(PRIN2010-2011, grant n. 2010NHBSBE).
The research leading to these results has received funding from the European
Commission Seventh Framework Programme (FP7/2007-2013) under grant agreement
n.267251 "Astronomy Fellowships in Italy" (AstroFIt).
SM and FJC acknowledge financial support by the Spanish Ministry of Economy and Competitiveness through grants AYA2010-21490-C02-01 and AYA2012-31447 and from the ARCHES project, funded by the 7th Framework of the European Union (project n. 313146). AR acknowledges financial support by the
Spanish Ministry of Economy and Competitiveness through grant AYA2012-31447.
Based on data from the Wide-field Infrared Survey Explorer, which is a joint project of the University of California, Los Angeles, and the Jet Propulsion Laboratory/California Institute of Technology, funded by the National Aeronautics and Space Administration. Funding for the SDSS and SDSS-II has been provided by the Alfred P. Sloan Foundation, the Participating Institutions, the National Science Foundation, the U.S. Department of Energy, the National Aeronautics and Space Administration, the Japanese Monbukagakusho, the Max Planck Society, and the Higher Education Funding Council for England. The SDSS Web Site is http://www.sdss.org/. Based on observations made with the  Telescopio Nazionale Galileo -operated by the Centro Galileo Galilei- and the Gran Telescopio de Canarias installed in the Spanish Observatorio del Roque de los Muchachos of the Instituto de Astrofisica de Canarias, in the island of La Palma (Spain).
We thank the anonymous referee for useful comments that have improved the quality of the paper.

\label{lastpage}

\end{document}